# Chiral kagome superconductivity modulations with residual Fermi arcs in $KV_3Sb_5$ and $CsV_3Sb_5$

**Authors:** Hanbin Deng[1]*, Hailang Qin[2]*, Guowei Liu[1]*, Tianyu Yang[1]*, Ruiqing Fu[3]*, Zhongyi Zhang[4], Xianxin Wu[3]†, Zhiwei Wang[5,6]†, Youguo Shi[7,8,9]†, Jinjin Liu[5,6], Hongxiong Liu[7,8], Xiao-Yu Yan[1], Wei Song[1], Xitong Xu[10], Yuanyuan Zhao[2], Mingsheng Yi[11], Gang Xu[11], Hendrik Hohmann[12], Sofie Castro Holbæk[13], Matteo Dürrnagel[12,14], Sen Zhou[3], Guoqing Chang[15], Yugui Yao[5,6], Qianghua Wang[16], Zurab Guguchia[17], Titus Neupert[13], Ronny Thomale[12], Mark H. Fischer[13], Jia-Xin Yin[1,2]†

**Affiliations:**

[1]Department of Physics, Southern University of Science and Technology, Shenzhen, Guangdong, China.

[2]Quantum Science Center of Guangdong-Hong Kong-Macao Greater Bay Area (Guangdong), Shenzhen, China.

[3]CAS Key Laboratory of Theoretical Physics, Institute of Theoretical Physics, Chinese Academy of Sciences, Beijing 100190, China.

[4]Department of Physics, Hong Kong University of Science and Technology, Clear Water Bay, Hong Kong, China.

[5]Centre for Quantum Physics, Key Laboratory of Advanced Optoelectronic Quantum Architecture and Measurement (MOE), School of Physics, Beijing Institute of Technology, Beijing, China.

[6]Beijing Key Lab of Nanophotonics and Ultrafine Optoelectronic Systems, Beijing Institute of Technology, Beijing, China.

[7]Beijing National Laboratory for Condensed Matter Physics and Institute of Physics, Chinese Academy of Sciences, Beijing 100190, China.

[8]University of Chinese Academy of Sciences, Beijing 100049, China.

[9]Songshan Lake Materials Laboratory, Dongguan, Guangdong 523808, China.

[10]Anhui Key Laboratory of Condensed Matter Physics at Extreme Conditions, High Magnetic Field Laboratory, Hefei Institutes of Physical Science, Chinese Academy of Sciences, Hefei, Anhui 230031, China.

[11]Wuhan National High Magnetic Field Center and School of Physics, Huazhong University of Science and Technology, Wuhan, China.

[12]Institute for Theoretical Physics and Astrophysics, University of Wurzburg, 97074 Wurzburg, Germany.

[13]Department of Physics, University of Zurich, Winterthurerstrasse, Zurich, Switzerland.

[14]Institute for Theoretical Physics, ETH Zürich, 8093 Zürich, Switzerland.

[15]Division of Physics and Applied Physics, School of Physical and Mathematical Sciences, Nanyang Technological University, Singapore 637371, Singapore.

[16]National Laboratory of Solid State Microstructures & School of Physics, Nanjing University, Nanjing 210093, China.

[17]Laboratory for Muon Spin Spectroscopy, Paul Scherrer Institute, CH-5232, Villigen PSI, Switzerland.

*These authors contributed equally to this work.

†Corresponding authors. E-mail:
xxwu@itp.ac.cn; zhiweiwang@bit.edu.cn; ygshi@iphy.ac.cn; yinjx@sustech.edu.cn

**Superconductivity involving finite momentum pairing[1] can lead to spatial gap and pair density modulations, as well as Bogoliubov Fermi states within the superconducting gap. However, the experimental realization of their intertwined relations has been challenging. Here, we detect chiral kagome superconductivity modulations with residual Fermi arcs in $KV_3Sb_5$ and $CsV_3Sb_5$ by normal and Josephson scanning tunneling microscopy down to 30mK with resolved electronic energy difference at µeV level. We observe a U-shaped superconducting gap with flat residual in-gap states. This gap exhibits chiral 2×2 spatial modulations with magnetic field tunable chirality, which align with the chiral 2×2 pair density modulations observed through Josephson tunneling. These findings demonstrate a chiral pair density wave (PDW) that breaks time-reversal symmetry. Quasiparticle interference imaging of the in-gap zero-energy states reveals segmented arcs, with high-temperature data linking them to parts of the reconstructed V $d$-orbital states within the charge order. The detected residual Fermi arcs can be explained by the partial suppression of these $d$-orbital states through an interorbital 2×2 PDW and thus serve as candidate Bogoliubov Fermi states. Additionally, we differentiate the observed PDW order from impurity-induced gap modulations. Our observations not only uncover a chiral PDW order with orbital-selectivity, but also illuminate the fundamental space-momentum correspondence inherent in finite momentum paired superconductivity.**

The kagome lattice, a periodic pattern of corner-sharing triangles, naturally hosts Dirac fermions, flat bands, and van Hove singularities in its electronic structure. The Dirac fermions encode topology, flat bands favor correlated ferromagnetism, and van Hove singularities can lead to instabilities toward long-range many-body orders. These electronic features have led to widespread interest in kagome lattice materials[2-5]. Among the kagome materials, a highly unexpected observation concerns the unconventional charge order occurring at bulk 2×2 ordering vectors in kagome superconductors $AV_3Sb_5$ (A=K, Rb, Cs), featuring unusual phenomena intertwining electronic chirality, charge density wave, and time-reversal symmetry-breaking[6-21]. These rich kagome charge order phenomena bridge the fields of topological matter and electron correlation. However, the interplay between the charge order and superconductivity in the ground state remains elusive in both experiment and theory. In this work, we further advance the frontier of this research direction by observing chiral superconductivity modulations with Bogoliubov Fermi states, utilizing dilution-refrigerator-based normal and Josephson scanning tunneling microscopy.

**Charge-ordered kagome superconductivity**

We focus on the charge-ordered kagome superconductor[22] $KV_3Sb_5$, as it is free from the extra stripy charge order[23-25] that exists on the surfaces of $RbV_3Sb_5$ and $CsV_3Sb_5$ but is yet to be confirmed by bulk scattering techniques including neutron and X-ray scattering[26]. Its low superconducting transition temperature[22] $T_C$ = 0.93K challenges the spectroscopic characterization of its superconducting properties, which have not yet been reported. Given the low $T_C$, our scanning tunneling microscopy measurements are mainly performed at a lattice temperature of 30mK, unless otherwise specified. Figure 1**a** shows an atomically resolved topographic image of a large clean Sb surface. The Sb layer is tightly bound to the V-kagome layer (Fig.1**b**), and is one natural cleavage surface. The Fourier transform of the topographic image demonstrates the 2×2 charge order (Fig. 1**c**). The charge order leads to a gap-like feature[7,27] near the Fermi level $E_F$ with an energy range of ±20meV (Fig. 1**d**). In the low energy range of ±0.5meV, we detect an additional U-shaped gap with a pair of coherence peaks at

±0.17meV (Fig. 1**e**). As this gap disappears above $T_C$, we identify it as the superconducting gap. Including the effect of the charge gap, the flat residual states in the superconducting phase are less than 10% of the total normal state density of states at $E_F$. We apply a magnetic field of 10mT along the *c*-axis, and detect a zero-energy vortex core state by mapping over a large area (Figs. 1**f, g**). The detection of the vortex core state further confirms our identification of the pairing gap. The pairing condensate of the sample is also supported by our tunneling experiment with a superconducting tip (Nb), showing a much larger total pairing gap (Fig. 1**h**). We detect a zero-bias peak in the differential conductance by reducing the tip-sample distance and the characteristic double kink features in the related current-voltage spectrum, both of which serve as key signatures of the Josephson tunneling signal and indicate the Cooper pair tunneling between two superconductors[24,28,29]. These experiments provide spectroscopic evidence for charge-ordered kagome superconductivity.

**2*a* pairing gap and pair density modulations**
The interplay between charge order and superconductivity in a kagome lattice can be highly intriguing, and there are proposals for different kinds of finite momentum pairing states with PDW orders[24,30-34]. To search for a PDW state[28,29,35-38], we perform high-resolution measurements of the pairing gap and pair density modulation in an impurity-free region. We first measure the low-energy differential conductance g($r$, $E$) along a line on the clean Sb surface shown in Fig. 2**a**. From the fine measurement of the coherence peaks in Fig. 2**b**, we detect modulations on their energy positions. We extract the energy positions of the coherence peaks at both negative bias $\Delta_{SC}^-$ and positive bias $\Delta_{SC}^+$. Then, we obtain the modulation of the pairing gap $\Delta_{SC}(r) = [\Delta_{SC}^+(r) - \Delta_{SC}^-(r)]/2$ in the left panel of Fig. 2**c**, which can be well described by a cosine function with 2*a* periodicity. The detection of 2*a* pairing gap modulation encourages us to measure the spatial modulation of the Josephson tunneling signal.

The scanned Josephson tunneling signal[29] can measure the spatial modulation of the phenomenologically defined Cooper pair density $N_J(r)$, which is another way to search for a PDW state. While point Josephson tunneling spectroscopy has been achieved in a kagome superconductor[24], the Josephson scanning tunneling microscopy has yet to be realized owing to the small pairing gap and complex surface environment of the cleaved kagome superconductors. Here, we overcome these challenges and demonstrate the atomically resolved Josephson tunneling signal. It has been shown[29] that $N_J(r) \propto g_J(r, E = 0) \times R_N^2(r)$, where $g_J(r, E = 0)$ is the spatially resolved zero-bias peak in the differential conductance data and $R_N(r)$ is the spatially resolved Josephson junction resistance at which the current-voltage characteristics is linear (see Methods for more details). We implement this method to detect the pair density modulation, and our scanned Josephson zero-bias-peak data with atomic resolution in Figs. 2**d** and **e** show that $N_J(r)$ can be well described by a cosine function with the 2*a* periodicity. The 2*a* modulation of the pairing gap, coherence peak height, and pair density is further revealed by their Fourier transformation analysis in Figs. 2**f-h**. Therefore, the pairing gap modulation and pair density modulation unambiguously establish the existence of the 2*a* PDW.

**Chiral 2×2 pairing gap and pair density modulations**

One decisive way to detect pairing modulations in two dimensions is by measuring the gap map $\Delta_{SC}(r)$

in the impurity-free region, which is presented in Figs. 3**a-b** as well as Extended Data Fig. 2. The Fourier transform of the gap map resolves three pairs of 2×2 vector peaks (Fig. 3**b** inset). Similar to the chiral charge order[7-10], the intensities of the 2×2 vector peaks are different, which defines a clockwise chirality when counting from low to high (see Extended Data Fig. 3 for the difference between chirality and nematicity). The inverse Fourier transform of the 2×2 vector peaks is shown in Fig. 3**c**, from which one can see that the 2*a* pairing gap modulations are different along the three *a*-axes, elaborating the chirality in real space. Using the Josephson scanning tunneling microscopy, we also map the pair density $N_J(r)$ in the impurity-free region in Figs. 3**d** and **e**. The Fourier analysis again resolves chiral 2×2 pair density modulations (Figs. 3**e** inset and **f**). The robust chiral 2×2 pairing gap and pair density modulations together establish a chiral 2×2 PDW.

The chiral PDW is intimately related to the charge order that exhibits magnetic field switchable chirality as demonstrated in Extended Data Fig. 4. Intriguingly, the chirality of the charge order gap modulation is reversed compared with the chirality of the pairing gap modulation at the same region (Extended Data Figs. 5 and 6), highlighting an unusual coupling between charge order and PDW. In Extended Data Fig. 7, we also find that the chirality of this PDW is switchable by the magnetic field. An initial understanding of the magnetic field switch of the chiral charge modulations is to consider a winding phase between the triple-**Q** order parameters of the charge order[7]. It is conceivable that the chiral 2×2 PDW can also carry such a winding phase, which breaks time-reversal symmetry. Our observation of chiral PDW at the bulk 2×2 ordering vectors is also in line with the transport detection[39-42] of the anisotropic Josephson effect, charge 6e superconductivity, and superconducting diode effect, all requiring breaking composite symmetries in the superconducting state.

**Residual Fermi arcs under the orbital selective PDW**
While we have resolved the PDW order, we now turn our attention to the residual in-gap states (Fig. 1**e**). The existence of these states is consistent with previous thermal conductivity and muon spin resonance data indicative of substantial in-gap excitations[43,44]. The flat gap bottom is highly unusual, which distinguishes it from a nodal pairing state. We uncover the residual states' momentum geometry by measuring the quasi-particle interference[45] (QPI) at $E_F$ from 30K to 30mK in Fig. 4**a-h** (see also Extended Data Fig. 8). We symmetrize the data to enhance its signal-to-noise ratio, thus omitting chirality. The QPI signal at high temperatures mainly occurs at a circular vector $q_p$ that is from the scattering of the Sb *p*-orbital pocket. Below $T_C$, $q_p$ is substantially suppressed (Figs. 4**f-h**), indicative of uniform superconductivity occurring on the *p*-orbital. Meanwhile, reducing the temperature from 4.2K to 600mK, a regular triangle shape signal $q_d$ progressively emerges (Fig. 4**d-f**), which is consistent with scattering between the 2×2 reconstructed *d*-orbital pockets within the charge order (Fig. 4**i**). This signal is not clearly detected at higher temperatures probably owing to the lack of coherence in the correlative *d*-orbital and a significant background signal at $q_p$. At 30mK, comparable to the PDW gap energy scale, the triangle signal evolves into well-defined segmented arc-like signals $q_{d1}$ and $q_{d2}$ (Fig. 4**h**). The arc-like signals at base temperature provide clear momentum signatures of the residual in-gap states, to which we consequently refer as residual Fermi arcs.

The detection of residual Fermi arcs first suggests the strong reduction of uniform pairing in the *d*-orbital derived states, which can be related to the time-reversal symmetry-breaking feature of the chiral charge order developed therein. The momentum geometry of the residual Fermi arcs is explained by considering the orbital structure of the PDW order, which can occur in the *d-d*, *d-p* or *p-p* channel. Our theoretical analysis suggests that the *d-p* PDW order can produce arc-like Bogoliubov Fermi states (see Supplementary Information for details), a finding that aligns more closely with experimental observation. The physical process is schematically illustrated in Figs. 4**i-k**: as the 2×2 charge order gaps out van Hove singularities of V *d*-orbital and introduces in-plane band folding, the remaining *d*-orbital triangles can intersect with the *p*-orbital pocket; the corresponding finite momentum (1/2×1/2) *p-d* pairing can open a PDW gap at their Fermi surface crossing points (Fig. 4**j**) and the Bogoliubov particle-hole mixing enhances the intensities of the inner corners of the *d*-orbital triangles; upon considering the uniform superconductivity in the *p*-orbital, which gaps out the circular pockets, we are left with arc-like Bogoliubov Fermi states (Fig. 4**k**), whose scattering can generate experimentally observed QPI signals. The 3D charge order can further fold the circular pocket with a larger size from $k_z+\pi$ and the *d-p* PDW will generate similar arc-like features. The uniform superconductivity develops mainly in the Sb *p*-orbital likely owing to a phonon-mediated pairing, while the charge order develops mainly in the V *d*-orbital likely owing to electron-electron correlations, and the coupling of these two orders occurring in distinct orbitals produces an interorbital PDW, which involves Cooper pairs with momenta $-\boldsymbol{k}_p$ and $+\boldsymbol{k}_d+\boldsymbol{Q}_{PDW}$.

**Impurity-induced gap modulations**

For completeness, we also measure the gap map for the defect-rich region and perform a Fourier transform shown in Fig. 4**l** (see also Extended Data Fig. 9), which shows remarkable pairing modulations at extra vectors $\boldsymbol{q}_{d1}$, $\boldsymbol{q}_{d2}$ and 3/(4a)×3/(4a), as compared with the gap map of the defect-free region (Fig. 3). Our observation of extra pairing gap modulations is consistent with the proposed impurity pair-breaking scattering interference under broken time-reversal symmetry[46]. This is also consistent with the muon spin resonance data[13] that the time-reversal symmetry breaking develops readily at the charge-ordered state and persists into the superconducting state. Our systematic data clarifies that the pairing gap modulations at 3/(4a)×3/(4a) vectors[24,47] as unusual impurity pair-breaking scattering interferences[46] between the tiny pockets reconstructed from the extra triangular Fermi surfaces centered around K points[47]. Our distinction of the PDW at the impurity-free region with impurity-induced pair modulation is fundamental for the theoretical modeling of PDW in kagome superconductors, and provides an important experimental reference for distinguishing the PDW order in other materials.

**Conclusions**

Our high-resolution and systematic tunneling data provide rich information on the ground state of the kagome superconductor, which links to a wide range of unusual phenomena of kagome charge order and superconductivity[2-5]. We build the space-momentum correspondence of the finite momentum pairing, as well as distinguish PDW order with impurity-induced pairing modulation. We show that both $KV_3Sb_5$ and $CsV_3Sb_5$ exhibit chiral 2×2 PDW order as well as residual Fermi arcs, while $CsV_3Sb_5$ features additional pairing modulation channels related to the (surface) stripes as detailed in Extended Data Fig. 10. Our key observations are summarized in Fig. 5: we detect both pair gap and pair density modulations at the bulk charge ordering vector of the kagome lattice (Fig. 5**a**); their 2×2 modulations exhibit magnetic field switchable chirality, underlining a chiral PDW order with broken time-reversal symmetry (Fig. 5**b**); in correspondence to the symmetry of this PDW order, we detect residual Fermi arcs of the V *d*-orbital states in the QPI data as candidate Bogoliubov Fermi states (Fig. 5**c**). The space-momentum correspondence of finite-momentum pairing elaborates the chiral PDW to be in the *p-d* interorbital channel. Beyond kagome systems, our systematic spectroscopic methods at extreme conditions pave the way for elucidating intertwined orders in superconductors with $T_C$ below 1K.

kagome lattice with a fractional vortex-antivortex pairing transition. Phys. Rev. B 108, 014424 (2023).
43. Guguchia, Z. et al. Tunable unconventional kagome superconductivity in charge ordered $RbV_3Sb_5$ and $KV_3Sb_5$. Nat Commun 14, 153 (2023).
44. Zhao, C. C. et al. Nodal superconductivity and superconducting domes in the topological Kagome metal $CsV_3Sb_5$. Preprint at arXiv:2102.08356 (2021).
45. Zhu, Z. et al. Discovery of segmented Fermi surface induced by Cooper pair momentum. Science 374, 1381-1385 (2021).
46. Gao, Z.-Q., Lin, Y.-P. & Lee, D.-H. Pair-breaking scattering interference as a mechanism for superconducting gap modulation. Preprint at arXiv 2310.06024 (2023).
47. Li, H. et al. Small Fermi pockets intertwined with charge stripes and pair density wave order in a kagome Superconductor. Phys. Rev. X 13, 031030 (2023).**Figure legends**

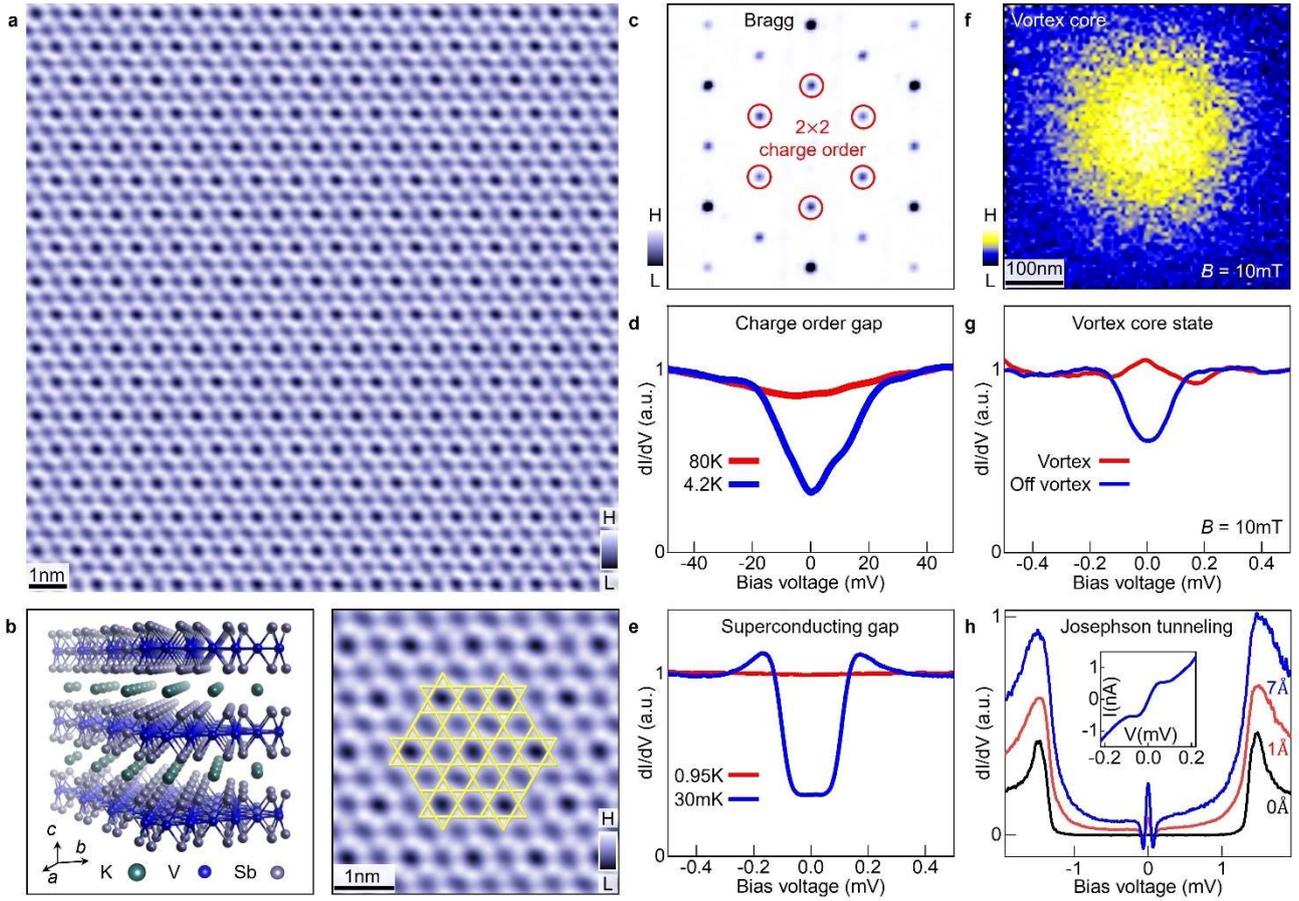

**Figure. 1 Charge-ordered kagome superconductivity. a,** Topographic image of a clean Sb surface, showing the 2×2 lattice modulation. Each Sb atom is resolved. **b,** The left panel shows the crystal structure of $KV_3Sb_5$. The right panel shows a zoom-in image of the Sb surface. The yellow lines on top illustrate the underlying kagome lattice. **c,** Fourier transform of the topographic image, showing three pairs of 2×2 vector peaks as marked by the inner red circles. **d,** A differential conductance spectrum at 4.2K shows the charge order gap opened near $E_F$. This gap disappears at the charge order transition temperature at 80K. **e,** Differential conductance spectrum (blue) showing the superconducting gap with a pair of coherence peaks. The red curve shows the disappearance of the superconducting gap when elevating the temperature to above $T_C$ ($T_C$ = 0.93K). **f,** Differential conductance map taken at zero-energy with applying a magnetic field of 10mT along *c*-axis, showing the emergence of a vortex core state. **g,** Differential conductance taken at and off the vortex core, showing a zero-energy vortex core state. **h,** Differential conductance taken with reducing the sample tip distance (from 0Å to -7Å in reference to normal tunneling condition) using a superconducting tip, showing the emergence of Josephson tunneling signal at zero bias. The inset shows the typical I-V curve of the Josephson tunneling signal.

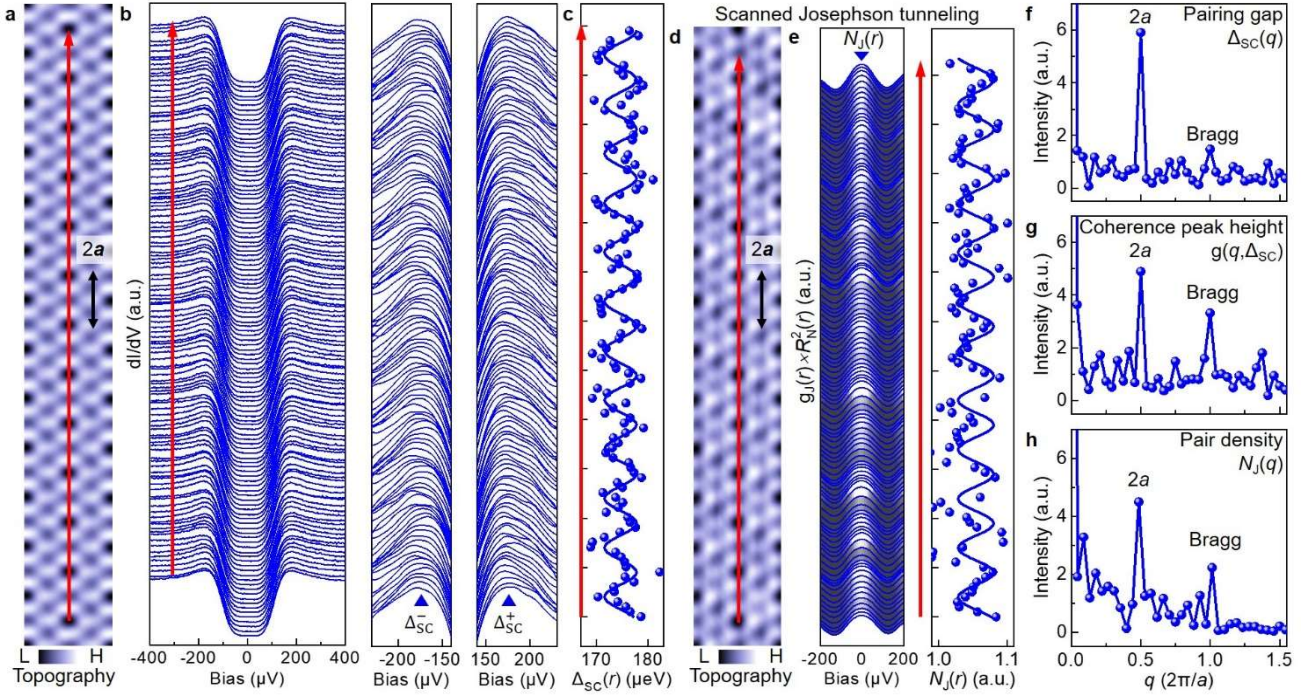

**Figure 2. 2*a* pairing gap and pair density modulations. a,** Atomically resolved Sb surface. The red line shows the positions where the differential conductance spectra are taken. **b,** Differential conductance spectra taken to resolve the superconducting gap modulation. The right panels show higher resolution spectra near the superconducting coherence peaks at ±170μeV, from which we determine out $\Delta_{SC}^{\pm}(r)$. **c,** Pairing gap modulations $\Delta_{SC}^{\square}(r) = [\Delta_{SC}^{+}(r) - \Delta_{SC}^{-}(r)]/2$, and the blue line shows a cosine function fitting with a 2*a* periodicity. **d,** Atomically resolved Sb surface with the superconducting tip. The red line shows the positions where the differential conductance spectra are taken. **e,** Differential conductance spectra $g_J(r, E)$ under scanned Josephson tunneling times a coefficient $R_N^2$. The right panel shows the spatially resolved pair density $N_J(r) \propto g_J(r, E = 0) \times R_N^2(r)$, and the blue line shows a cosine function fitting with a 2*a* periodicity. **f,** Fourier transform of the superconducting gap $\Delta_{SC}^{\square}(q)$, showing the 2*a* modulation. **g,** Fourier transform of the coherence peak height $g(q, \Delta_{SC}^{\square})$, showing the 2*a* modulation. The coherence peak height modulation is calculated as $g(r, \Delta_{SC}^{\square}) = [g(r, \Delta_{SC}^{-}) + g(r, \Delta_{SC}^{+})]/2$, and $g(q, \Delta_{SC}^{\square})$ is the corresponding Fourier transform. **h,** Fourier transform of the pair density $N_J(q)$, showing the 2*a* modulation.

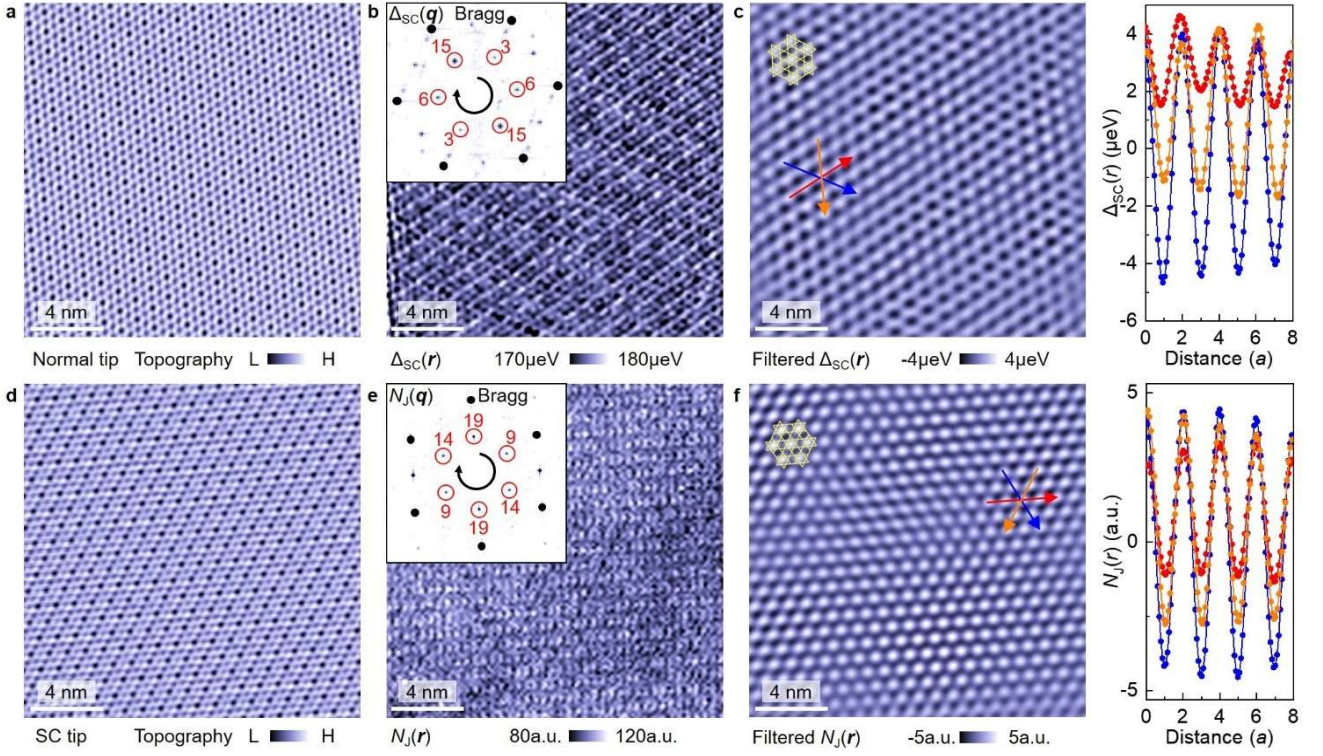

**Figure 3. Chiral 2×2 pairing gap and pair density modulations. a,** Topographic image of Sb surface in $KV_3Sb_5$. **b,** Corresponding pairing gap map. The inset is its Fourier transform, showing 2×2 vector peaks with different intensities. The intensity is a direct readout of the peak value. Counting from lower to higher intensity peaks, we determine its chirality to be clockwise. **c,** Inverse Fourier transform of the 2×2 vector peaks. Yellow lines mark the underlying kagome lattice. The right panel shows three line profiles along the lines marked in the left panel, which demonstrates the anisotropic modulation strength along three different directions as the illustration of chirality in real space. **d,** Topographic image of Sb surface in $KV_3Sb_5$ measured with a superconducting tip. **e,** Corresponding pair density map detected by Josephson tunneling microscopy. The inset is its Fourier transform, showing 2×2 vector peaks with different intensities defining clockwise chirality. **f,** Inverse Fourier transform of the 2×2 vector peaks, and the yellow lines mark the underlying kagome lattice. The right panel shows three line profiles along the lines marked in the left panel, which demonstrates the anisotropic modulation strength along three different directions as the illustration of chirality in real space.

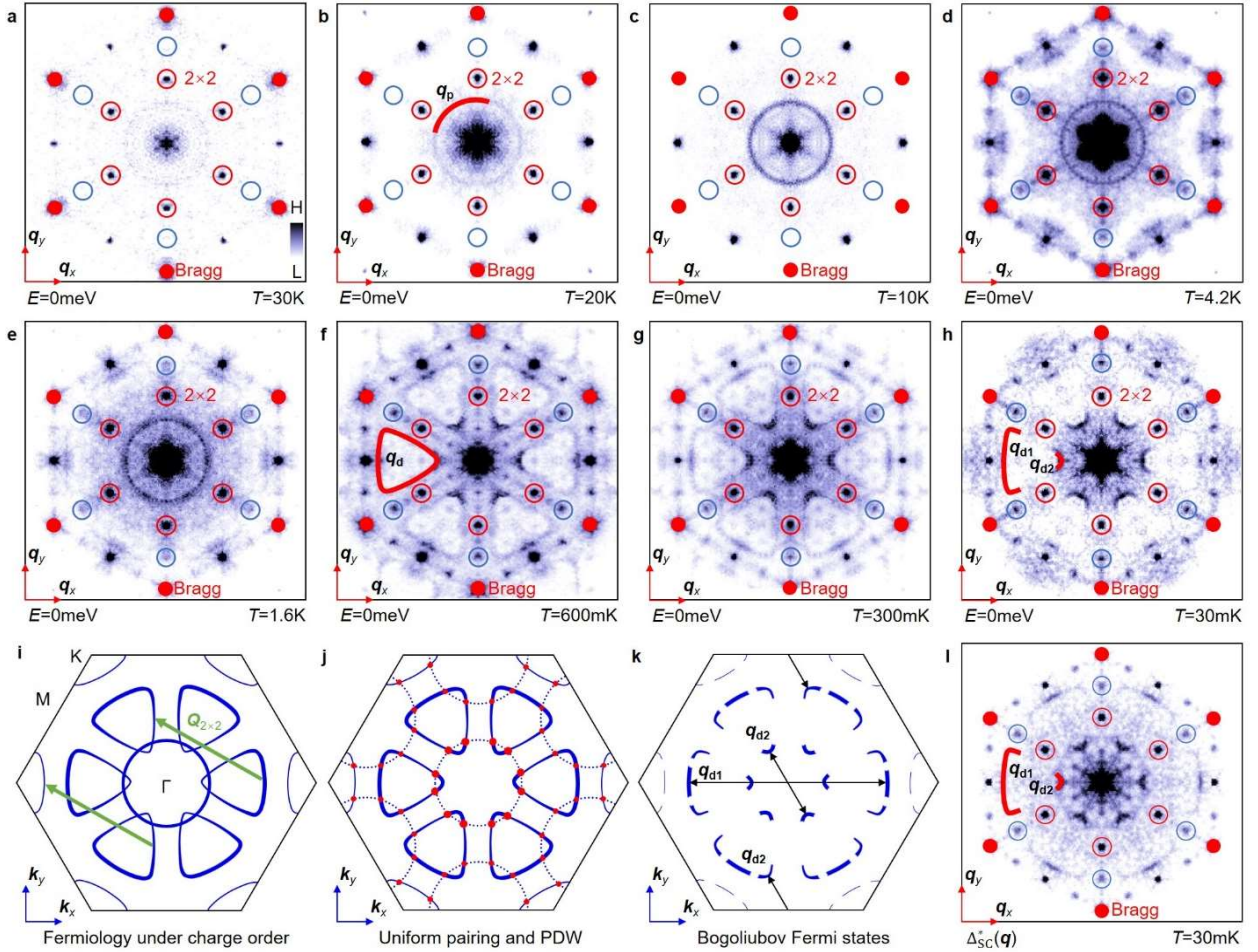

**Figure 4. Residual Fermi arcs under orbital-selectivity. a, b, c, d, e, f, g, h,** Symmetrized QPI at zero-energy from 30K to 30mK. Red circles mark 2×2 vector, and blue circles mark the 3/(4*a*) vector. The red curves in **b** mark its leading interference signals at $q_p$, which is consistent with the scattering of the circular pocket with Sb *p*-orbital character. Below $T_C$, $q_p$ is substantially suppressed (**f-h**), indicative of uniform superconductivity occurring on the *p*-orbital. Meanwhile, reducing the temperature from 4.2K to 600mK, a regular triangle shape signal $q_d$ progressively emerges (**f**), which is consistent with scattering between the reconstructed *d*-orbital pockets under 2×2 band folding. The 30mK data (**h**) eventually shows the emergence of arc-like signals $q_{d1}$ and $q_{d2}$. **i,** Illustration of the reconstructed Fermi surfaces within the charge order that gaps out the van Hove singularities of V *d*-orbital near M and leads to segmented regular triangles. The green arrows mark the folding vector $\mathbf{Q}_{2\times2}$. **j,** The uniform superconductivity occurs in the *p*-orbital circular pocket (dash lines), which can intersect with the reconstructed *d*-orbital triangle pockets. The corresponding finite momentum pairing opens a PDW gap at their crossing points (marked by red dots), and the Bogoliubov particle-hole mixing enhances the intensities of the inner corners of the *d*-orbital triangles. Owing to the 3D nature of the charge order, we anticipate the folded triangle pockets mainly interact with the slightly larger circular pocket at $k_z+\pi$. **k,** Expected arc-like Bogoliubov Fermi states under a *p-d* PDW and a uniform pairing on the *p*-orbital. The scattering of the Bogoliubov Fermi states can lead to arc-like QPI signals at $q_{d1}$ and $q_{d2}$. **l,** Symmetrized Fourier transform of the pairing gap map at the impurity-rich region, showing extra impurity-induced gap modulations at $q_{d1}$ and $q_{d2}$ and 3/(4*a*) as compared with the defect-free region in Fig.3. Red circles mark 2×2 vector peaks.

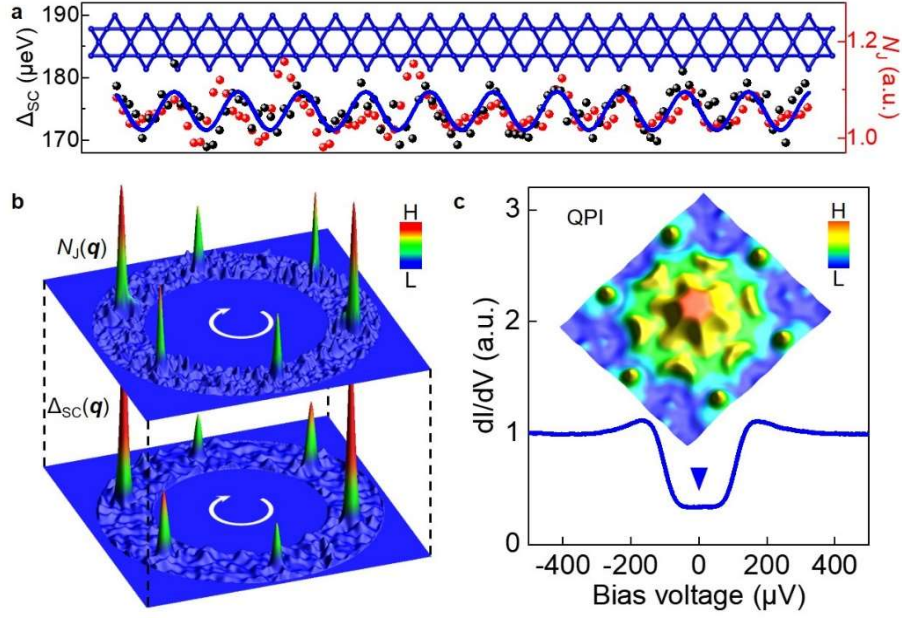

**Figure 5. Evidence chain for finite momentum paired superconductivity. a,** 2*a* pairing gap (black dots) and pair density (red dots) modulations extracted from Fig. 2. The blue curve represents a cosine function with 2*a* periodicity. The inset shows the kagome lattice reference. **b**, 2×2 pairing gap and pair density modulations with chirality marked by the white arrow. The 2×2 vector peaks are extracted from Fig. 3. **c,** Evidence for residual Fermi arcs from in-gap differential conductance states (main panel, extracted from Fig. 1) and arc-like QPI data at zero-energy (inset, extracted from Fig. 4). The QPI data are symmetrized and we only show the essential part within the 2×2 vector peaks.

## Methods
### Scanning tunneling microscopy at extreme conditions

Single crystals are grown with the same methods described in Ref. 22. Crystals with sizes up to 2mm×2mm×0.5mm are cleaved mechanically *in situ* at 20K in ultra-high vacuum conditions, and then immediately inserted into the microscope head, already at He4 base temperature (4.2K). More than 50 crystals are cleaved and imaged in this project. We then further cool the microscope head to 30mK via a dilution refrigerator. Tunneling conductance spectra are obtained with Ir/Pt tips (or a superconducting tip as described below) using standard lock-in amplifier techniques. We measure the superconducting gap of a related full gap kagome superconductor single crystal 14%-Ta doped $CsV_3Sb_5$ to estimate the electronic temperature of our system. By fitting the superconducting gap with the Bardeen-Cooper-Schrieffer gap function, we estimate the electronic temperature to be 90mK.

We extensively scan each crystal for large and clean Sb surfaces, which can take up to one week. Topographic images are usually taken with a tunneling junction set up of $V = 30$mV $I = 0.5$nA. The conductance maps and gap map are obtained by taking a spectrum at each location (with the feedback loop off) with tunneling junction set up: $V = 0.5$mV (For $KV_3Sb_5$), $V = 1$mV (For $CsV_3Sb_5$), $I = 1$nA, and modulation voltage $V_m$=3~20μV. The point spectrums were taken with the same condition. The magnetic field is applied under a zero-field cool condition. Overall, we find that the experimental challenge is not only the instrumentational realization of extremely low temperatures, but is essentially about the stable tunneling and scanning with extremely low junction resistances. Since we are measuring a small energy gap, the tunneling voltage is often set to be very low ~0.5mV; meanwhile, to obtain high energy resolution, the modulation voltage is often set to be extremely small ~5μV; lastly, with such low modulation, to obtain a large enough signal-to-noise ratio, the tunneling current is often set to be high ~1nA; eventually, they together determine that the junction resistance has to be very low. This tunneling condition further requires a clean sample area and very stable tunneling tips with atomic resolution. One way to prepare such a stable tip is to anneal the tip on an Au crystal with a similar tunneling condition to make the tip very robust.

**Josephson scanning tunneling microscopy**

In our scanned Josephson tunneling experiment[28,29,48-53], we use the Nb polycrystalline tip whose pairing gap is estimated to be 1.33meV judging from the total gap size of the Josephson junction. The superconducting tips are made from 0.2 mm diameter Nb wire by mechanical sharpening. We further anneal the tip with an e-beam heater to 1500°C to remove possible oxide layers on the tip[54]. We precheck the superconducting nature (pairing gap) on a Pb sample at 30mK, which produces a total gap of 2.65meV that is much larger than the measured gap on the kagome superconductor or the pairing gap of Pb. Owing to the complex surface environment (Sb surface mixed with K surfaces and adatoms and other disordered surfaces), it is extremely challenging to maintain the superconductivity (pairing gap) of the tip with atomic resolutions. More than 10 Nb tips are tried on the kagome superconductors. Topographic images are usually taken with a tunneling junction set up of $V$ = 30mV $I$ = 0.5nA. The Josephson conductance spectra and maps are obtained by taking a spectrum at each location (with the feedback loop off) with a typical tunneling junction setup: $V$ = 3mV, $I$ = 2~8 nA, and modulation voltage $V_m$=30μV. The zero-bias peak is readily evident at this junction setup (Fig. 2, 3), and is further enhanced by reducing the tip-sample distances (current up to 80nA) as in Fig. 1. The Josephson coupling energy $E_J$ is estimated to be 20mK·$k_B$ ($k_B$ is the Boltzmann constant), which is smaller than the electronic temperature of the system 90mK, thus making the Josephson junction close to the phase diffusive limit.

In principle, the phenomenologically defined local pair density $N_J(r)$ is expected to be proportional to the Josephson critical current $I_J(r)$ times the normal state resistance[55] $R_N(r)$: $N_J(r) \propto I_J^2(r) R_N^2(r)$. However, for small gap superconductors where the Josephson coupling energy $E_J$ is typically smaller than the electronic temperature, it is hard to accurately measure $I_J$, and the tunneling current takes the form: $I = 0.5 I_J^2 ZV/(V^2 + V_C^2)$, where $Z$ is the high-frequency impedance and $V_C$ is a characteristic voltage where the maximum phase-diffusive electron-pair tunneling current occurs. Then the differential conductance at zero-energy is $g_J(0) = dI/dV = 0.5 I_J^2 V_C^{-2} \propto I_J^2$. Therefore, $g_J(r, E = 0) R_N^2(r)$ is a practical way to measure the pair density $N_J(r)$ for small gap superconductors[29]. In the experiment, for a fixed location, we also find a linear relationship between $g_J(0)$ and $R_N^{-2}$ with different junction resistances (Extended Data Fig 1). This scaling behavior is highly expected for a Josephson junction and attests to the fact that our extracted local pair density $N_J(r)$ does not depend on the junction resistance setup. We, therefore, measure both zero-bias conductance peak $g_J(r, E = 0)$ and normal-state junction resistance $R_N(r)$ at a given location $r$. The $R_N(r)$ is taken as $V/I(r, V)$, where $V$ is the sample bias at which the current-voltage characteristics is linear (typically at 3mV) and $I(r, V)$ is tunneling current at bias V. We also note that if $R_N(r)$ is substantially larger than that of the lead resistance $R_{Lead}$ (resistance from the wiring, filter, etc.), $R_{Lead}$ can be ignored; however, if it is substantially reduced, $R_{Lead}$ has to be subtracted from $R_N(r)$ and $g_J(r, 0)$ also has to be recalibrated as $g = g_{measured}/(1 - g_{measured} \cdot R_{Lead})$ for better accuracy. We have measured $N_J(r)$ with both constant current mode and constant height mode, and the two methods yield consistent results.

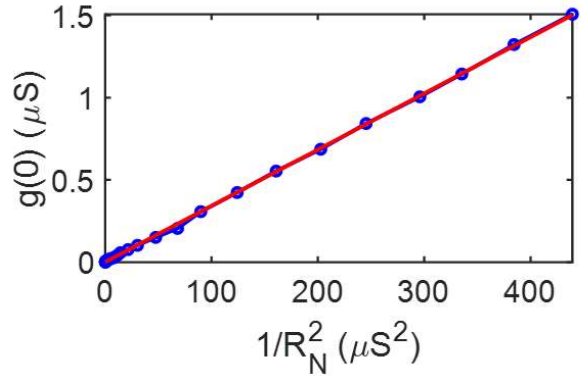

**Extended Data Fig. 1 Linear scaling relation between g(0) and $R_N^{-2}$ for Josephson tunneling.**

**Particle-hole symmetrical pairing gap modulation.**
Extended Data Fig. 2 shows the real space gap map for KV$_3$Sb$_5$ and CsV$_3$Sb$_5$ at 30mK. We take the gap map for positive bias $\Delta_{SC}^+(r)$ and negative bias $\Delta_{SC}^+(r)$, and demonstrate that they resemble each other, following the particle-hole symmetry for superconductivity.

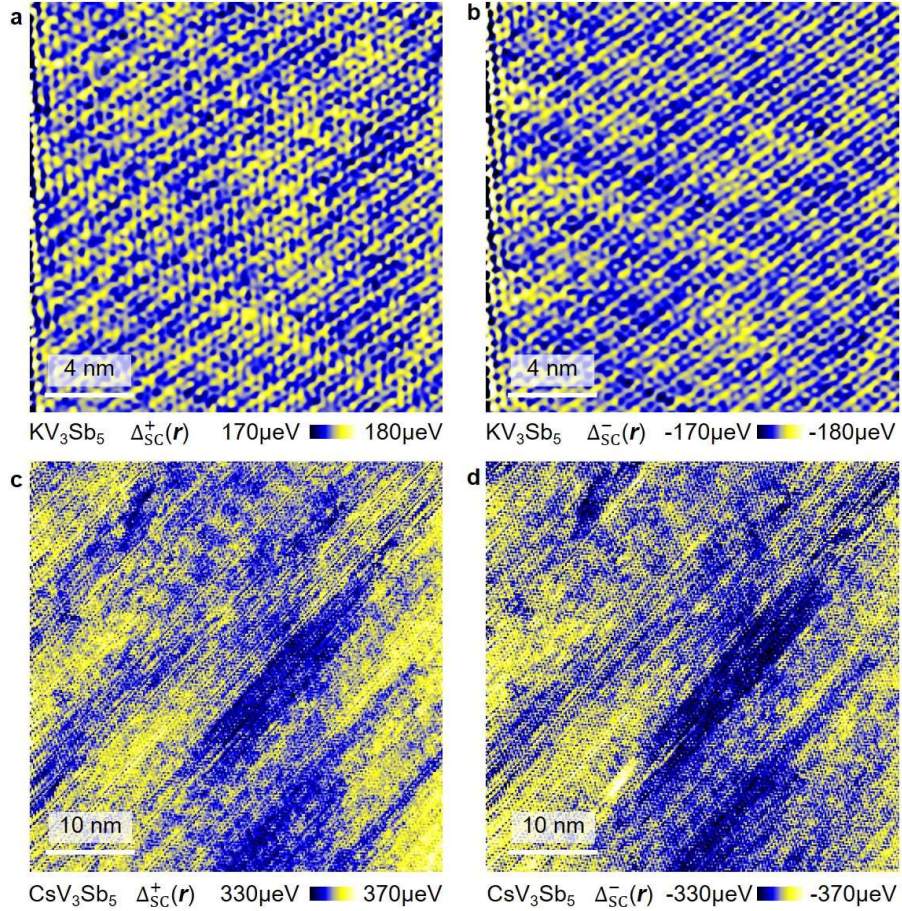

**Extended data Fig. 2 Particle-hole symmetry of pairing gap modulations for $KV_3Sb_5$ (a, b) and $CsV_3Sb_5$ (c, d).**

**Quasi-particle interference at defect-few (chiral) and defect-rich (nematic) regions at 4.2K**

We first note that to make the symmetry-breaking analysis of the 2×2 charge order precise, a clean imaging area larger than 200Å × 200Å is often required. The 2×2 charge order detected in the Sb surface in $AV_3Sb_5$ has been shown to be chiral in defect-free region[7-10] and nematic in defect-rich region[7,8,9,11,47]. It was also argued that even the clean region also showed the absence of chirality[11]. But a comparison of the related data on charge order gap and quasi-particle interference signals suggests a substantial deviation from the clean region and a possibility that the claimed small clean region is actually surrounded by defects (evidenced by detectable defect-induced quasi-particle interferences and shrinkage of the charge order gap as demonstrated in the detailed analysis in the supplementary of Ref. 56). The intrinsic nature of chirality of the charge order is supported by various experiments including magnetic field control of chirality[7,8,10], optical control of chirality[10], time-reversal symmetry-breaking[13,14], chiral transport[15], and anomalous Hall effect[6,9]. In line with these discussions, our zero-energy map data in Extended Data Fig. 3 also confirms that the chirality of 2×2 charge order emerges for defect-few regions and nematicity for defect-rich regions. In addition, we do not detect strong $4a/3$ signals for defect-few regions, and observe such signals only for defect-rich regions. This observation is consistent with previous studies suggesting they are not static orders but dispersive and originate from a defect-assisted scattering of smaller Fermi pockets[47]. It is worthwhile to note that in all the data that exhibit chirality (i.e. three 2×2 vector peaks have different intensities), we often observe that two of the three have close intensities, thus consistent with the coexistence of nematicity. Since the discovery of the 2×2×2 charge order in $AV_3Sb_5$, it has been pointed out that the phase shift between adjacent 2×2 charge orders lead to geometrical nematicity[19,57]. Such geometrical effect can lead to the nematic background of the 2×2 vector peak intensities.

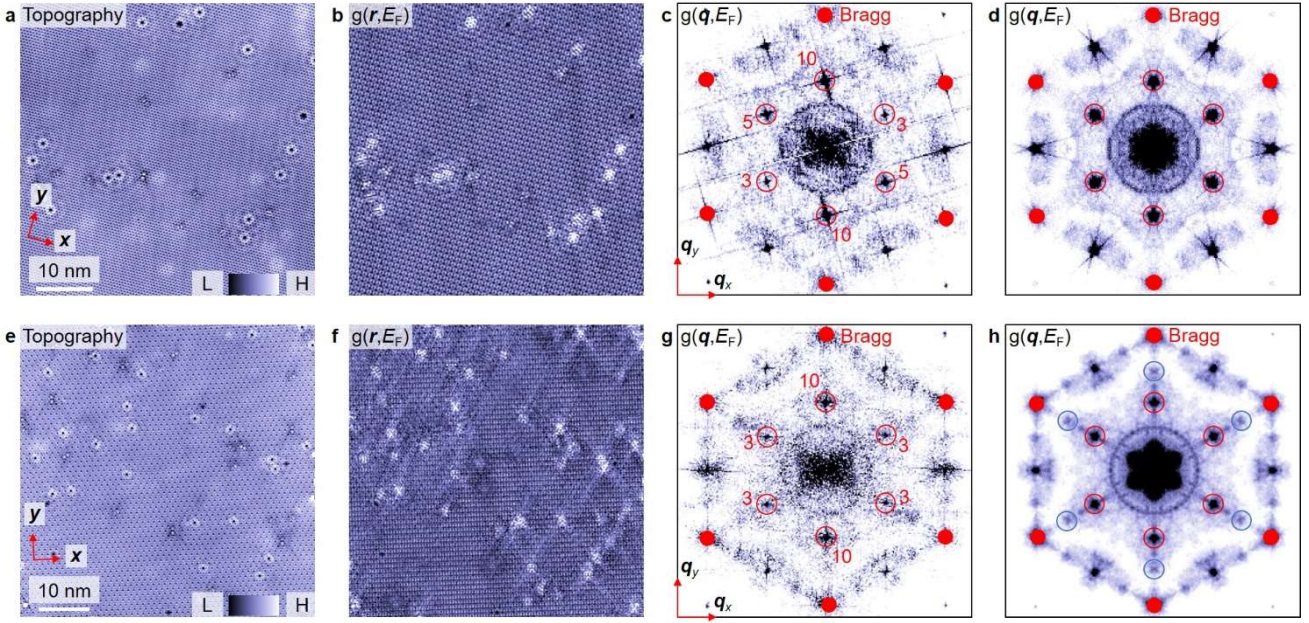

**Extended Data Fig. 3 Quasi-particle interference at defect-few (a-d) and defect-rich (e-h) regions at 4.2K. a,** A topographic image for the defect-few region. **b,** Corresponding zero-energy map. **c,** Fourier transform of the zero-energy map, showing chiral 2×2 vector peaks. The numbers mark the intensity of the 2×2 vector peaks. **d,** Six-fold symmetrization of the quasi-particle interference signal. **e,** A topographic image for the defect-rich region. **f,** Corresponding zero-energy map. **g,** Fourier transform of the zero-energy map, showing nematic 2×2 vector peaks. The numbers mark the intensity of the 2×2 vector peaks. **h,** Six-fold symmetrization of the quasi-particle interference signal, showing additional 3/(4a) vector peaks marked by blue circles as compared with **d**. All the data are taken at 4.2K.

**Chirality domains and field-induced chirality switch below the original $T_C$**

In the main text, we have shown that the 2×2 pairing gap modulations exhibit robust chirality, in analogy to the chiral charge order. In the superconducting state ($T$=300mK), we further demonstrate the magnetic field switch of the 2×2 vector peaks from the dI/dV intensity map at $E_F$. Extended Data Fig. 4 illustrates two chirality domains at 0T, separated by a dark line in the middle. Zooming into the clean region in I and II, the Fourier transform of the real space data shows the opposite chirality of the 2×2 vector peaks (Extended Data Figs. 4**b** and **c**). We further choose region II and apply a magnetic field of +2T along the $c$-axis. Note that although we are at the temperature below the original $T_C$, superconductivity is readily destroyed by the applied large magnetic field. The Fourier transform of the spectroscopic map data at $E_F$ shows clockwise chirality of the 2×2 vector peaks (Extended Data Fig. 4**d**). We further apply a field of -2T, and observe a switch of the chirality to anticlockwise (Extended Data Fig. 4**e**). The magnetic field induced chirality switch of the 2×2 vector peaks below the original $T_C$ is consistent with the time-reversal symmetry-breaking of the ground state of the system.

We further find that after applying the magnetic field, the domain wall disappears in the original field of view (Extended Data Fig. 4**f**).

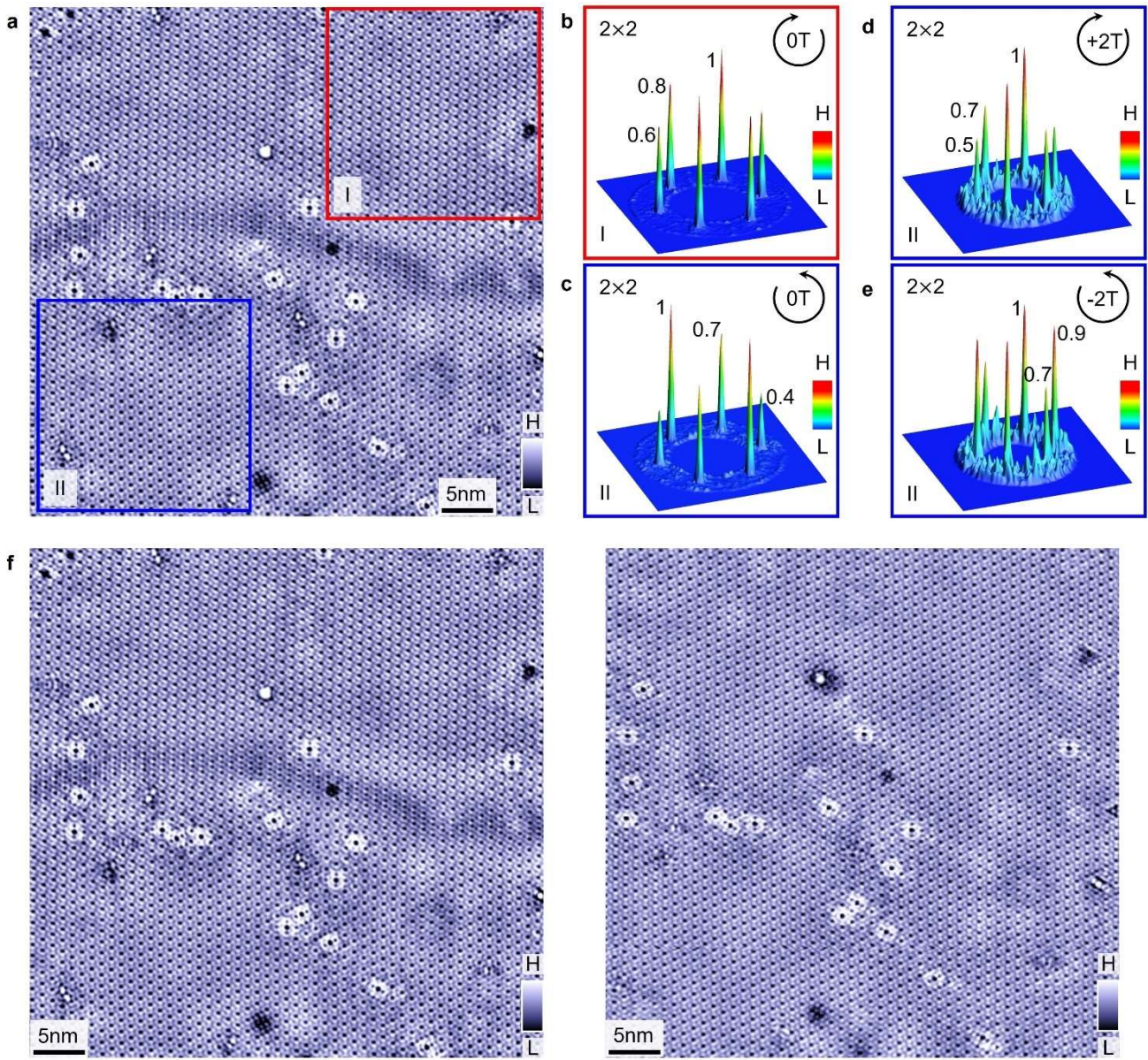

**Extended Data Fig. 4 Chirality domains and field-induced chirality switch below the original $T_C$.**
**a,** Topographic image for a large field of view, containing a domain wall as a dark line in the middle. The red and blue squares mark two clean regions as Region I and Region II respectively. **b,** 2×2 vector peaks of the Fourier transform of the topographic data in Region I at 0T showing clockwise chirality. **c,** 2×2 vector peaks of the Fourier transform of the topographic data in Region II at 0T showing anticlockwise chirality. **d,** 2×2 vector peaks of the Fourier transform of the zero-energy dI/dV map data in Region II at +2T showing clockwise chirality. **e,** 2×2 vector peaks of the Fourier transform of the zero-energy dI/dV map data in Region II at -2T showing anticlockwise chirality. Data was taken at 300mK, which is below the original $T_C$ of 0.93K. **f,** Disappearance of domain wall before (left panel) and after (right panel) the application of magnetic fields. Both data are imaged with the same tunneling junction setup at zero magnetic field. All data were taken at 300mK, which is below the original $T_C$ of

0.93K.

**Reversed chirality between charge order and PDW**

We elucidate the charge order gap modulation in Extended Data Fig. 5 at 30mK. We measure a series of differential conductance spectrums g($E$, $r$) (Extended Data Fig. 5**b**) along a line drawn in Extended Data Fig. 5**a**, which shows 2$a$ modulations. To enhance the visibility of this modulation, we performed curvature analysis[58] (Extended Data Figs. 5**c** and **d**). The curvature analysis along the distance direction highlights the 2$a$ charge modulations. Near $E_F$, the charge modulation reverses, hinting at a charge order gap $\Delta_{CO}$. The curvature analysis along the energy direction helps to extract the charge order gap at both negative bias $\Delta_{CO}^-$ and positive bias $\Delta_{CO}^+$. Then, we obtain the modulation of the charge order gap $\Delta_{CO}^{\square}(r) = \Delta_{CO}^+(r) - \Delta_{CO}^-(r)$ in the right panel of Extended Data Fig. 5**d**, which can be well described by a cosine function with a 2$a$ periodicity. In Extended Data Fig. 6, we show the charge order gap map and pairing gap map measured for the same atomic location, as well as their respective Fourier transform (Extended Data Figs. 6**a-e**). Intriguingly, the chirality of the 2×2 vector peaks of the charge gap is opposite to that of the pairing gap, suggesting an anti-chiral relation between the chiral charge order and chiral PDW. We further elaborate on their chirality in real space in Extended Data Figs. 6**f-i**. We find that the strongest modulation direction is the same in the real space, and the chirality switch is related to the intensity switch of the other two directions between the charge order and PDW.

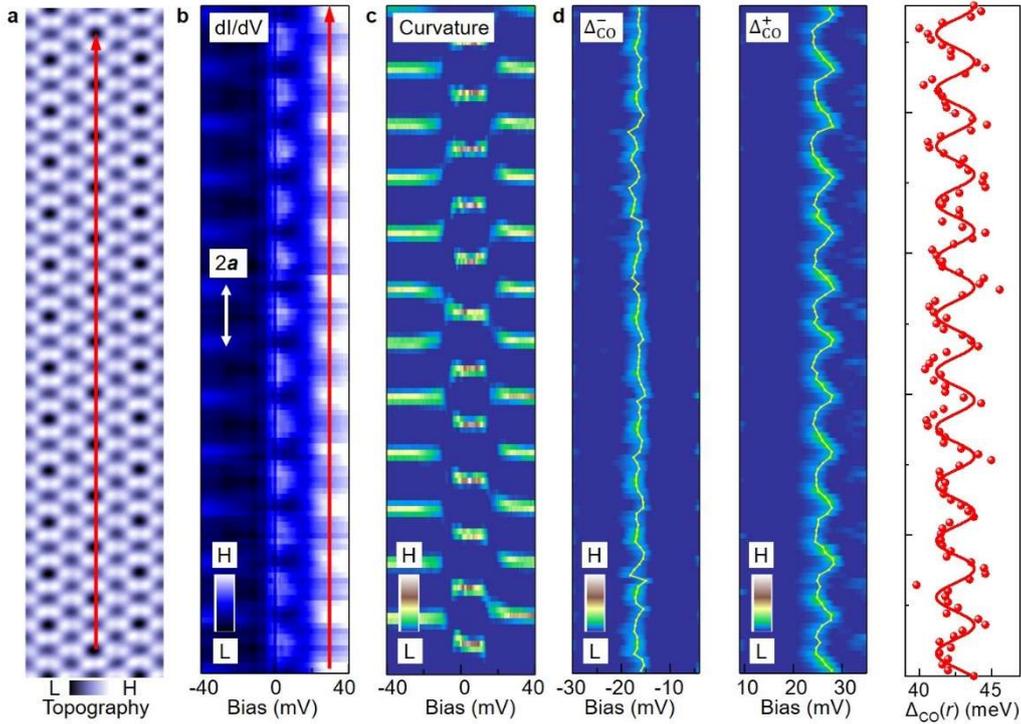

**Extended Data Fig. 5 Charge order gap modulations. a,** Atomically resolved Sb surface. The red line shows the positions where the differential conductance spectrums are taken. **b,** Intensity plot of the differential conductance spectrums taken to resolve the charge modulation and charge order gap modulation. **c,** A vertical (along the distance axis) curvature analysis of the line spectrums, showing the 2$a$ charge modulations. The charge modulation has an apparent reversal near the $E_F$, indicative of the existence of a charge order gap. **d,** A horizontal (along the energy axis) curvature analysis of the line spectrums, showing the charge order gap opens at $\Delta_{CO}^-$ of ~-18meV (left) and $\Delta_{CO}^+$ of ~+25meV (right). The yellow lines trace their local maximum value. The right panel shows the spatial modulation of the charge order gap $\Delta_{CO}^{\square}(r) = \Delta_{CO}^+(r) - \Delta_{CO}^-(r)$, where the red line shows a cosine function fitting with 2a periodicity.

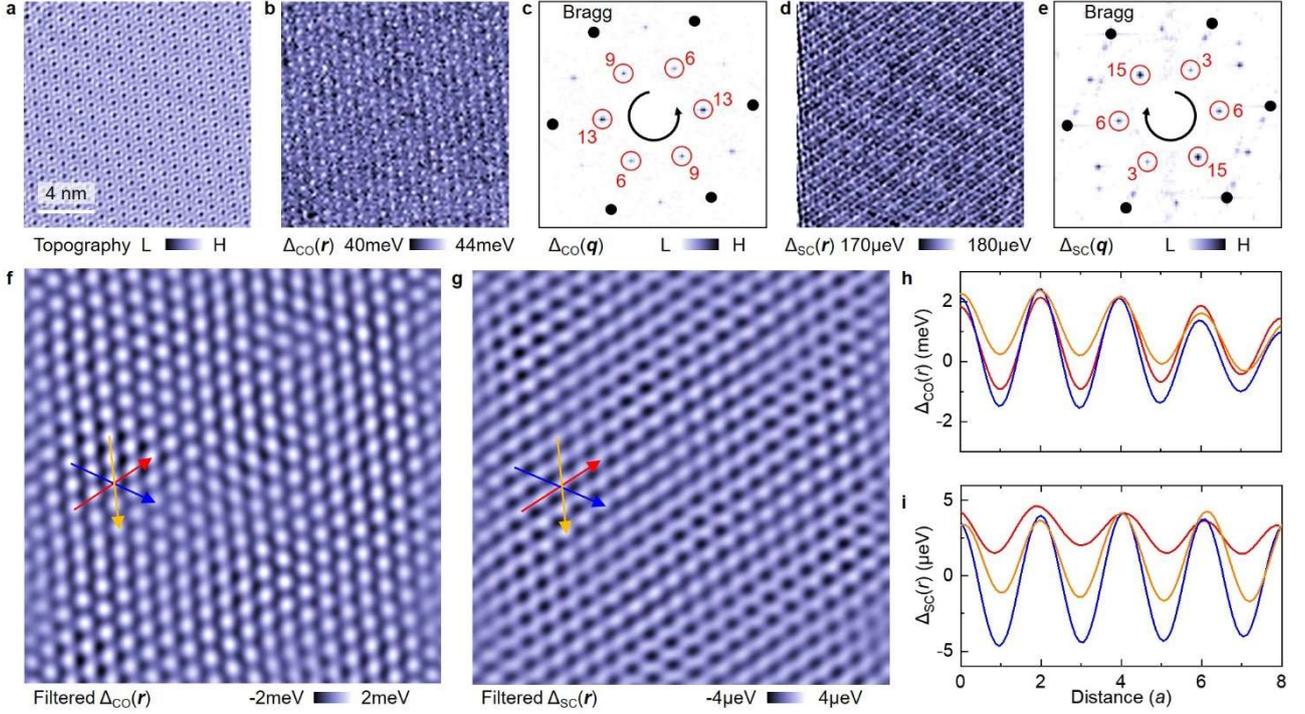

**Extended Data Fig. 6 Reversed chirality between charge order and PDW at the same location at 30mK. a,** Topographic image of a clean surface. **b,** Charge order gap map. **c,** Fourier transform of the charge order gap map, showing anticlockwise chirality of the 2×2 vector peaks. **d,** Pairing gap map. **e,** Fourier transform of pairing gap map, showing clockwise chirality of the 2×2 vector peaks. **f,** Inverse Fourier transform of the 2×2 vector peaks from the charge order gap. g, Inverse Fourier transform of the 2×2 vector peaks from the pairing gap. h, Three line profiles along the lines marked in **f**, which demonstrates the anisotropic modulation strength along three different directions as the illustration of chirality in real space. **i,** Three line profiles along the lines marked in **g**, which demonstrates the anisotropic modulation strength along three different directions as the illustration of chirality in real space.

**Chirality switch of the PDW order at 30mK**
We experimentally explore the chirality switch of the PDW order by magnetic field. Because the pairing is easily destroyed by a tiny external magnetic field, we have to measure at 0T with different field histories. We first measure the chirality of the pairing gap with zero-field cooling at 30mK (Extended Data Fig. 7**a**). We then apply the magnetic field of B=+1T along the *c*-axis, withdraw the field back to 0T, measure the chirality of the pairing gap at 30mK (Extended Data Fig. 7**b**), and find that the chirality of PDW order is switched. Finally, we apply the magnetic field of B=-1T along the *c*-axis, withdraw the field back to 0T, measure the chirality of the pairing gap at 30mK (Extended Data Fig. 7**c**), and find that the chirality of PDW order is switched back.

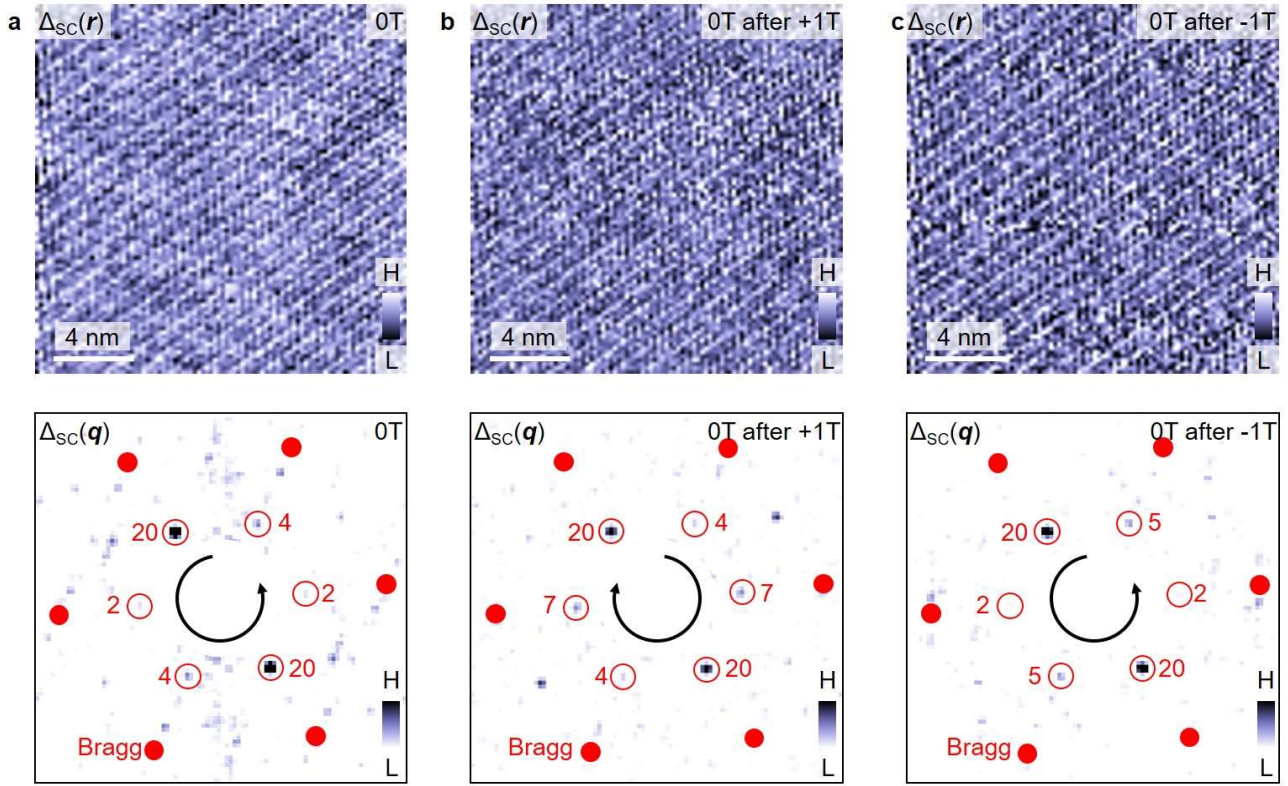

**Extended Data Fig. 7 Field history induced chirality switch of the PDW order at 30mK. a,** Pairing gap map (top) and its Fourier transform (bottom) showing anticlockwise chirality of the 2×2 vector peaks. **b,** Pairing gap map at 0T after +1T (top) and its Fourier transform (bottom) showing clockwise chirality of the 2×2 vector peaks. **c,** Pairing gap map at 0T after -1T (top) and its Fourier transform (bottom) showing anticlockwise chirality of the 2×2 vector peaks.

**Quasi-particle interference and gap map for the defect-rich region**

In Extended Data Fig. 8, we show the quasi-particle interference data taken from 30K to 30mK. In Extended Data Fig. 9, we show the zero-energy quasi-particle interference data and gap map data for the same area at 30mK as the extended data for Fig. 4 in the main paper. The inverse Fourier transform of the data at $q_1$, $q_2$, and $3/(4a)$ vectors show a positive correlation with the defects, demonstrating their intimate relation.

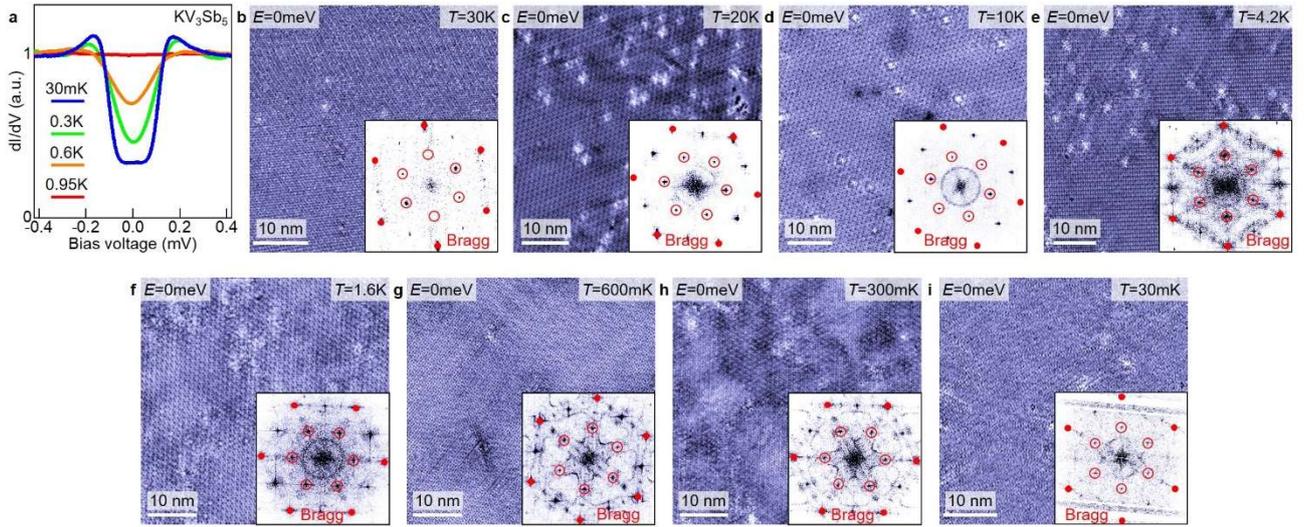

**Extended Data Fig. 8** Temperature-dependent tunneling spectrums below T$_C$ (a) and quasi-particle interference data from 30K to 30mK (b-i).

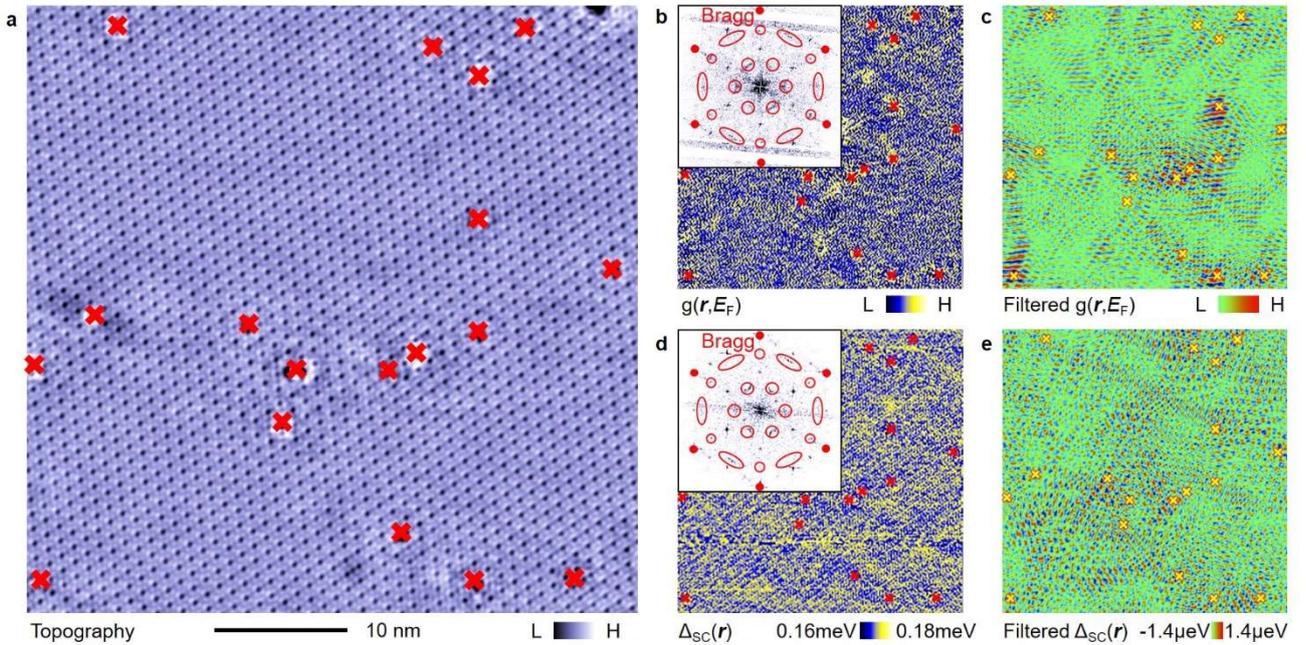

**Extended Data Fig. 9 Quasi-particle interference and gap map for defect region at 30mK. a,** Topographic image with defects. **b,** Corresponding zero-energy map. Defects are denoted with cross markers. The inset shows its Fourier transform, whose six-fold symmetrization data is shown in Fig. 4**h**. **c,** Inverse Fourier transform of the quasi-particle interference data in the inset of **b** marked with circles. Defects marked with cross markers have a positive correlation with the filtered signals. **d,** Corresponding gap map. Defects are denoted with cross markers. The inset shows its Fourier transform, whose six-fold symmetrization data is shown in Fig. 4**l**. **e,** Inverse Fourier transform of the gap modulation data in the inset of **d** marked with circles. Defects marked with cross markers have a positive correlation with the filtered signals.

## Chiral 2×2 PDW and residual Fermi arcs in CsV$_3$Sb$_5$

The tunneling spectrum of CsV$_3$Sb$_5$ at 30mK also shows a U-shaped pairing gap with flat in-gap states

(Extended Data Fig. 10a). Both the pairing gap map (Extended Data Figs. 10b and c) and pair density map (Extended Data Figs. 10d-f) show chiral 2×2 modulations, demonstrating a chiral 2×2 PDW order. We map the tunneling conductance at zero energy over a large field of view (Extended Data Figs. 10g and h), and its symmetrized Fourier transform (Extended Data Fig. 10i) also shows arc-like quasi-particle interference patterns, signaling residual Fermi arcs.

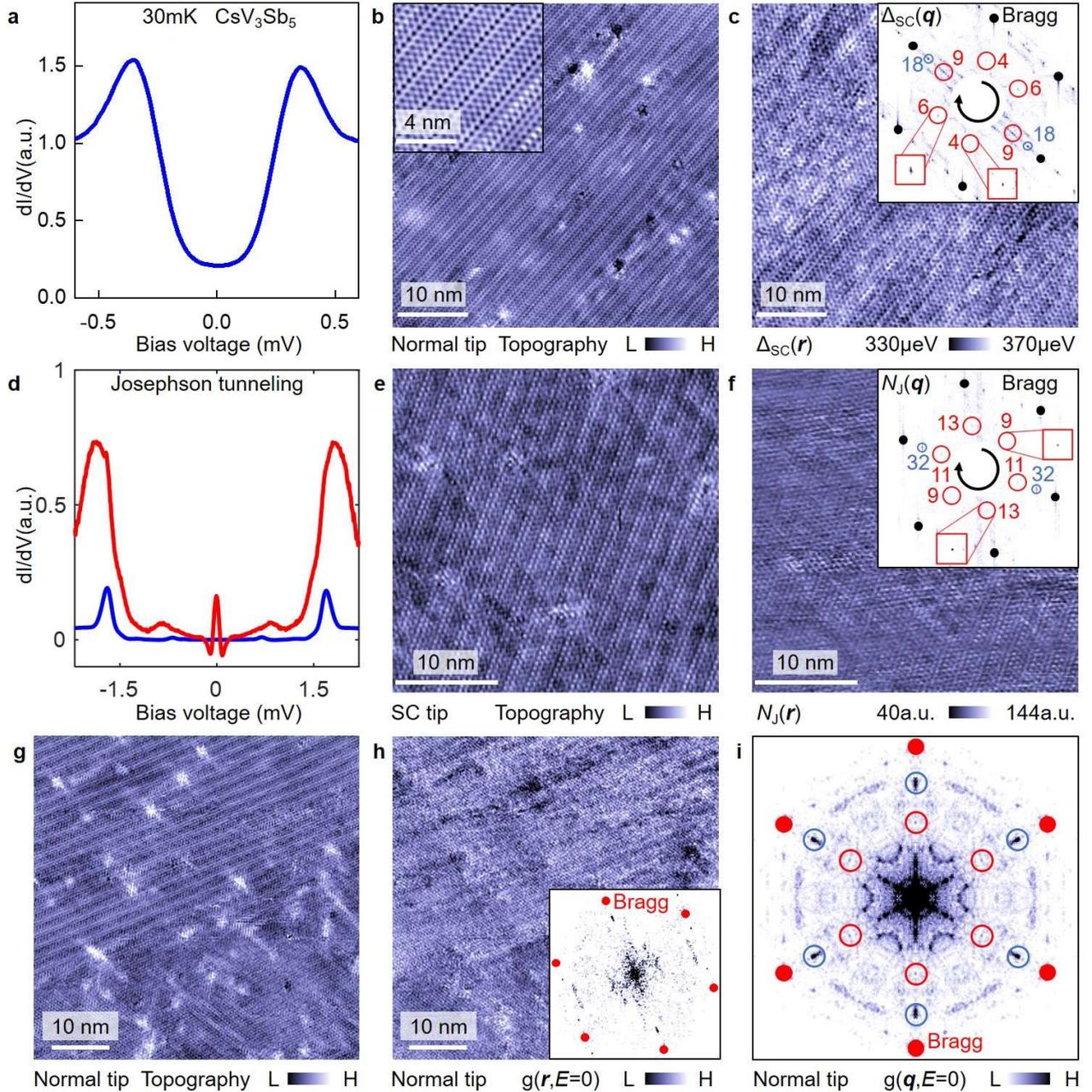

**Extended Data Fig. 10 Chiral 2×2 PDW and residual Fermi arcs in $CsV_3Sb_5$. a,** Tunneling spectrum taken at 30mK. **b,** Topographic image of the Sb surface. The inset shows a zoomed in image. **c,** Corresponding pairing gap map. The inset shows its Fourier transform, demonstrating 2×2 gap modulations (marked by red circles) with clockwise chirality and 1×4a/3 pairing gap modulation perpendicular to the stripe direction. **d,** Tunneling spectrums measured by reducing the sample tip distance from 0Å (blue curve) to -7Å (red curve) in reference to normal tunneling condition using a superconducting tip, showing the emergence of Josephson tunneling signal at zero bias. **e,** Topographic image measured with the superconducting tip. **f,** Corresponding air density map. The inset shows its Fourier transform, demonstrating 2×2 pair density modulations (marked by red circles) with clockwise chirality and 1×4a/3 pair density modulation perpendicular to the stripe direction. **g,** Topographic image. **h,** Corresponding zero-energy differential conductance map. The inset shows its Fourier transform. **i,** Symmetrized Fourier transform, showing arc-like quasi-particle interference features. All the data are taken at 30mK.

**Origin of pairing modulations at 3/(4$a$) vectors**

Our systematic data also provide insights into the pairing modulations at 3/(4$a$) vectors as previously identified[24] in $CsV_3Sb_5$. In ref. 47 as well as Ref. 30, it has been pointed out that the 3/(4$a$) vector connects tiny pockets near K points in the charge-ordered phase omitted in our main text. We now list several experimental facts that can help to further constrain its interpretation for the observation of this vector peak in the pairing gap map at the temperature well below $T_C$.

Firstly, in Extended Data Fig. 2 we show that the density of states modulation near 3/(4$a$) is absent at the defect-few region and appears at the defect-rich region at 4.2K, consistent with its interpretation[47] as originating from the impurity-assisted quasi-particle scattering of Fermi pockets' hot spots, rather than a spontaneous density wave order that will appear irrespective of impurities and will be nondispersive. Secondly, at defect-few regions of $CsV_3Sb_5$, the 3/(4$a$) pair modulation is only detected perpendicular to the stripes (Extended Data Fig. 10), which is consistent with previous research[47] at 4.2K showing that the density of states modulation near 3/(4$a$) is static (non-dispersive) only for this special direction. Thirdly, the 3/(4$a$) signal is absent for the gap map at the defect-free region at 30mK (Fig. 3), but appears in the defect-rich region in $KV_3Sb_5$ at 30mK (Fig. 4); the 3/(4$a$) signal is absent in the pair density map with scanned Josephson tunneling microscopy (Fig. 3). These pieces of evidence are consistent with the interpretation that the appearance of 3/(4$a$)×3/(4$a$) signal in the gap map of $KV_3Sb_5$ and $CsV_3Sb_5$ is an impurity pair-breaking scattering interference[46] and 1×3/(4$a$) in the gap map and pair density map of $CsV_3Sb_5$ might be a PDW related to the (surface) stripes.

**Electronic temperature estimation**

We measure the superconducting gap of a related full gap kagome superconductor single crystal 14%-Ta doped $CsV_3Sb_5$ to estimate the electronic temperature of our system. By fitting the superconducting gap with the BCS gap function (Extended Data Fig. 11), we estimate the electronic temperature to be around 90mK.

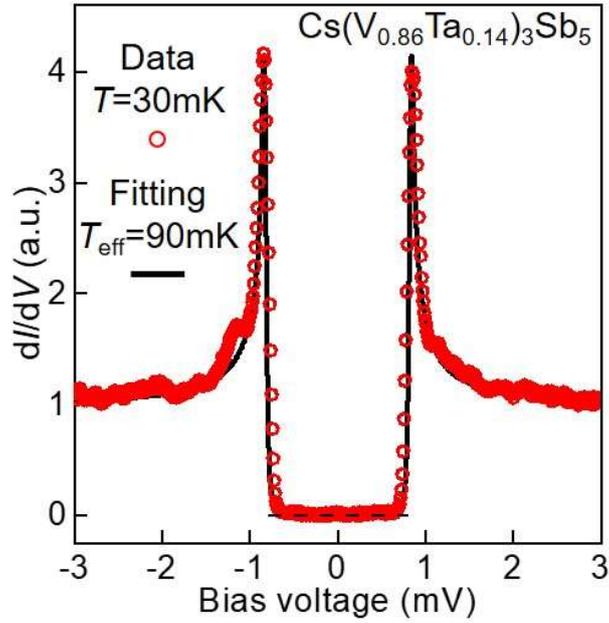

**Extended Data Fig. 11 Fitting the superconducting gap of Cs(V$_{0.86}$Ta$_{0.14}$)$_3$Sb$_5$ single crystal with BCS gap function.**

**Method references:**

**Acknowledgments**
We thank D. H. Lee, Z. Y. Weng, X. J. Zhou, Xi Dai, Xiaolong Liu and Guangming Zhang for insightful discussions. We thank M.Z. Hasan for an initial discussion on topological PDW. We thank S. H. Pan for sharing his experience in preparing superconducting tips with atomic resolution. We acknowledge the support from the National Key R&D Program of China (Nos. 2023YFA1407300, 2023YFF0718403, 2021YFA1401500, 2022YFA1403400, 2020YFA0308800, 2022YFA1403800), the National Science Foundation of China (Nos. 12374060, 12374147, 92365023, 12047503, 12374153, 12274154), and Guangdong Provincial Quantum Science Strategic Initiative (GDZX2201001). ZW is also supported by the Beijing Natural Science Foundation (Grant No. Z210006), and the Beijing National Laboratory for Condensed Matter Physics (Grant No. 2023BNLCMPKF007). MSY and GX acknowledge the support from the HPC Platform of Huazhong University of Science and Technology. Work at Nanyang Technological University was supported by the National Research Foundation, Singapore, under its Fellowship Award (NRF-NRFF13-2021-0010), the Agency for Science, Technology and Research (A*STAR) under its Manufacturing, Trade and Connectivity (MTC) Individual Research Grant (IRG) (No. M23M6c0100), Singapore Ministry of Education (MOE) AcRF Tier 2 grant (MOE-T2EP50222-0014) and the Nanyang Assistant Professorship grant (NTU-SUG). YY is supported by the NSFC (No. 12321004). HH, MD and RT received funding from the DeutscheForschungsgemeinschaft (DFG, German Research Foundation) through Project-ID 258499086 - SFB 1170 and the research unit QUAST, FOR 5249, project ID 449872909 and through the Würzburg-Dresden Cluster of Excellence on Complexity and Topology in Quantum Matter-ct.qmatProject-ID 390858490 - EXC 2147. ZG acknowledges support from the Swiss National Science Foundation (SNSF) through SNSF Starting Grant (No. TMSGI2_211750). MHF is supported by the Swiss National Science Foundation (SNSF) through Division II (No. 207908). TN acknowledges support from the Swiss National Science Foundation through a Consolidator Grant (iTQC, TMCG-2_213805). SCH was supported by the Swiss National Science Foundation (Project 200021E_198011) as part of the FOR 5249 (QUAST) led by the Deutsche Forschungsgemeinschaft (DFG, German Research Foundation).

**Competing interests**
The authors declare no competing interests.

**Author contributions**
H.D., H.Q., G.L., T.Y., X.Y.Y., and W. S. conducted the scanning tunneling microscopy experiments in consultation with J.X.Y.; R.F., Z.Z., X.W., Y.Z., H.H., S.C.H., M.D., S.Z., Y.Y., Q.W., T.N., R.T., and M.H.F. contributes to the theoretical understandings; Z.W., Y.S. J.L., H.L., X.X. and Z.G. synthesized and characterized the transport and susceptibility of samples; M.Y., G.X. and G.C. conducted first-principles calculations; H.D., H.Q., X.W. and J.X.Y. performed the data analysis and figure development and wrote the paper with contributions from all authors; J.X.Y. supervised the project.

**Correspondence and requests for materials** should be addressed to J.X.Y.

**Data availability**

All data are available in the main text or the supplementary materials.

**Supplementary Information**

**Theoretical analysis for Bogoliubov Fermi states**

Before we discuss a microscopic model of the origin of the Bogoliubov Fermi states, we present a short symmetry analysis as a guide. In particular, we consider the interplay of the charge order and superconductivity. For the charge order, the 2×2 modulation corresponds to the three momentum vectors to the M points in the Brillouin zone. For our analysis, we assume that the charge order is time-reversal-symmetry breaking and its energetics is governed by some Ginzburg-Landau functional $F_{CDW}[\chi_{Q_\alpha}]$ with the three inequivalent wave vectors $Q_\alpha$. Since the band around $\Gamma$ is fully gapped, we consider an order parameter $\Delta_0$ corresponding to a gap function and additionally introduce a PDW order parameter $\Delta_{Q_\alpha}$. The Ginzburg-Landau free energy for these additional order parameters to second order reads $F^{(2)}[\Delta_0, \Delta_{Q_\alpha}] = a_0(T)|\Delta_0|^2 + \sum_\alpha a_{PDW}(T)|\Delta_{Q_\alpha}|^2$, where a sign change in $a_i(T)$ signals a phase transition into the respective order. Additionally, there are third-order terms that couple the isotropic superconducting order, the charge order, and the PDW. Specifically, we have $F^{(3)}[\chi_{Q_\alpha}, \Delta_0, \Delta_{Q_\alpha}] = \sum_\alpha i\, b\, \chi_{Q_\alpha}(\Delta_0 \Delta_{Q_\alpha}^* - c.c.)$ with $b$ as a phenomenological coupling constant. This term mixes PDW and isotropic superconductivity in the presence of the charge order with the same wave vector $Q_\alpha$. Put differently, if one of the two orders is condensed, it induces the respective other order. Note that no conclusions regarding the microscopic details of the individual orders can be drawn from this Ginzburg-Landau analysis.

To analyze the situation in more detail and capture the main characteristics of the Fermi surface, we use a 4-band tight-binding model to capture the main characteristics of the Fermi surface. We limit our model to the two-dimensional lattice, thus the size of the Fermi pockets can be slightly different from the illustrative figures in Fig. 4 (that match the experiments considering three-dimensionality of the band folding). In this model, three orbitals come from the three kagome sublattice sites, and the other orbital is located at the center of hexagons (the $p_z$ orbital of Sb). The hopping parameters are provided in Table S1 and the obtained band structures and Fermi surfaces are displayed in Fig. S1.

| $t_1$ | $s_1$ | $p_1$ | $E_{p_z}$ | $\mu$ |
|---|---|---|---|---|
| -0.5 | 0.1 | -0.2877 | 1.2201 | 0.055 |

**Table S1. Tight-binding parameters.** $t_1$ is the nearest-neighbor hopping in the kagome lattice. $s_1$ is the hopping between nearest kagome and $p_z$ orbitals. $p_1$ is the nearest hopping between $p_z$ orbitals and $E_{pz}$ is the onsite energy of the $p_z$ orbital. $\mu$ is the overall chemical potential.

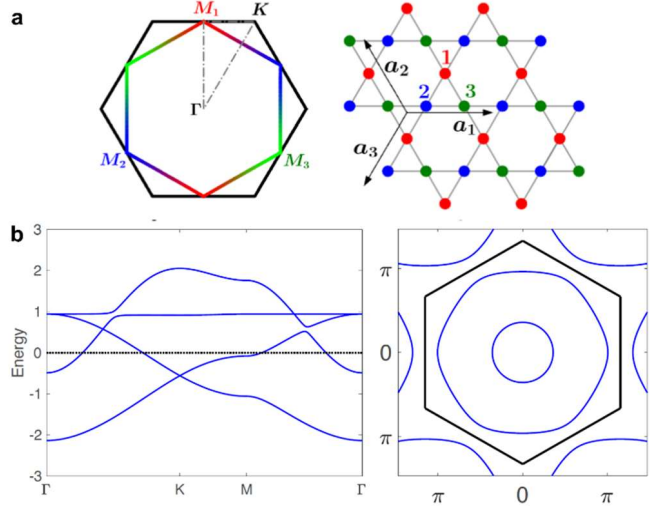

**Figure S1 Kagome lattice, band structure and Fermi surfaces. a,** The related Brillouin zone and kagome lattice. **b,** Normal state band structures and Fermi surface.

Now we further include the charge ordering in the above model. Due to the suggested time-reversal symmetry-breaking of the charge order[7-21], we consider the 2 × 2 imaginary charge bond order (loop current order) and our following analysis about Bogoliubov Fermi states can also be applied to the real charge bond order. The index 1-3 represent three kagome sublattices and 4 denotes the Sb $p_z$ orbital. The kagome lattice convention is displayed in Fig. S1**a**. We consider an anti-symmetric loop current order on the nearest-neighbor bonds in the kagome lattice and it reads,

$$\chi_\alpha = \frac{1}{2} \sum_{\beta,\gamma \in \{\epsilon_{\alpha\beta\gamma}=1\}} \epsilon_{\alpha\beta\gamma}\left(c^\dagger_{\beta,\mathbf{r}}c_{\gamma,\mathbf{r}+\mathbf{a}_\alpha/2} - c^\dagger_{\beta,\mathbf{r}}c_{\gamma,\mathbf{r}-\mathbf{a}_\alpha/2}\right)$$

where $\epsilon_{\alpha\beta\gamma}$ is the Levi-Civita symbol, $\alpha, \beta, \gamma \in 1,2,3$ and $\mathbf{a}_\alpha$ is defined in Fig. S1**a**. Here, we set this loop current order as $\langle\chi_\alpha(\mathbf{r})\rangle = \rho_0\cos(\mathbf{M}_\alpha \cdot \mathbf{r})$ with $\rho_0 = 0.05i$, where the vectors $\mathbf{M}_\alpha$ are given in the Fig. S1**a**. We illustrate the Fermiology under such charge ordering in Fig. S2.

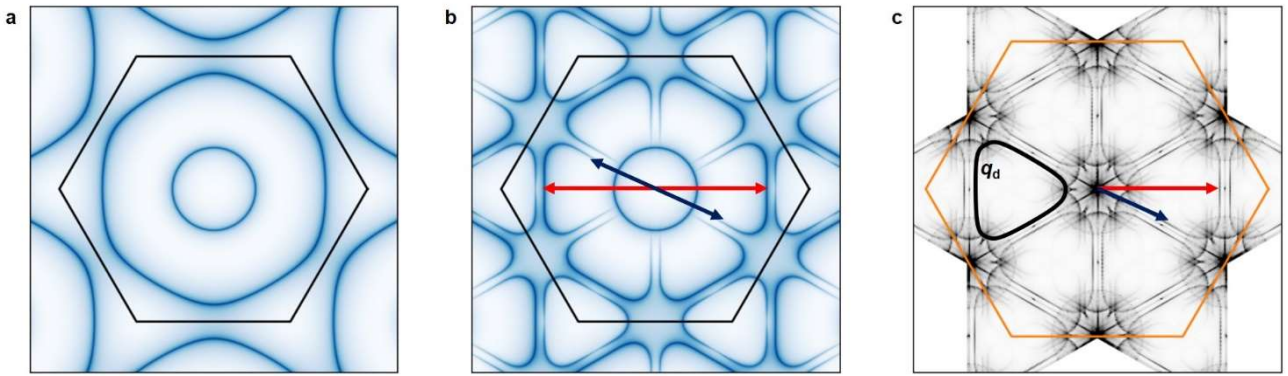

**Figure S2 Impact of charge order on the Fermiology. a,** Intensity plot of the original Fermi surface spectral weight. **b,** Fermi surface under charge ordering. **c,** Simulated quasi-particle interference pattern (joint density of states) of the Fermi surface under charge ordering. The triangle marks the experimentally detected triangle signal $q_d$. The signals marked by red and black arrows correspond to the scattering channels marked in **b**, respectively.

We further explore superconducting orders based on experimental observations. A U-shape gap is observed in our measurements, indicating a dominant $s$-wave pairing. Considering that the charge order gaps out a significant portion of the density of states from the kagome $d$ orbitals, and that these $d$ orbital states are present at zero energy while the circular pocket is notably absent in the quasi-particle interference measurements at zero energy at 30mK, we infer that the primary pairing $\Delta_0$ comes from the Sb $p_z$-orbital and the normal pairing in the V $d$-orbital almost vanishes. Moreover, these charge orders in $d$ orbitals and $s$-wave pairing in the $p$ orbital can induce a PDW order with the same modulation vector of charge order, i.e. $\Delta_{Q_\alpha}(\mathbf{r}) \propto \chi_{Q_\alpha}(\mathbf{r})\Delta_0^*(\mathbf{r})$. In principle, the PDW state can occur in the $d-d$, $d-p$, or $p-p$ channels. Including the charge order, $s$-wave pairing and the PDW, the full Hamiltonian reads

$$\begin{aligned}\mathcal{H}_T &= \mathcal{H}_0 + \mathcal{H}_{\text{CDW}} + \mathcal{H}_{\text{SC}} + \mathcal{H}_{\text{PDW}} \\ &= \mathcal{H}_{\text{PDW}} + \sum_{\mathbf{r}} \sum_{\alpha,\beta,\gamma \in \{\epsilon_{\alpha\beta\gamma}=1\}} [t_1 + \chi_\alpha(\mathbf{r}_\alpha)]\epsilon_{\alpha\beta\gamma} c^\dagger_{\beta,\mathbf{r}} c_{\gamma,\mathbf{r}+\mathbf{a}_\alpha/2} + [t_1 - \chi_\alpha(\mathbf{r}_\alpha)]\epsilon_{\alpha\beta\gamma} c^\dagger_{\beta,\mathbf{r}} c_{\gamma,\mathbf{r}-\mathbf{a}_\alpha/2} \\ &+ \sum_{\mathbf{r}} \sum_{\alpha} s_1(c^\dagger_{\alpha,\mathbf{r}} c_{p,\mathbf{r}+\mathbf{a}_\alpha/2} + c^\dagger_{\alpha,\mathbf{r}} c_{p,\mathbf{r}-\mathbf{a}_\alpha/2}) + p_1 c^\dagger_{p,\mathbf{r}} c_{p,\mathbf{r}+\mathbf{a}_\alpha} + (E_{p_z} - \mu)c^\dagger_{p,\mathbf{r}} c_{p,\mathbf{r}} - \mu c^\dagger_{\alpha,\mathbf{r}} c_{\alpha,\mathbf{r}} + [\Delta_0 c_{p,\mathbf{r}} c_{p,\mathbf{r}}\end{aligned}$$

In the following, we discuss the residual Bogoliubov Fermi surfaces in three types of PDW states and compare them with our measurements.

Firstly, we consider the $d-p$ PDW state and the corresponding Hamiltonian is,

$$\mathcal{H}^{dp}_{\text{PDW}} = \sum_{\mathbf{r}} \sum_{\alpha} \left[\Delta_{pd_\alpha}(\mathbf{r}_p)\left(c_{p,\mathbf{r}_p} c_{\alpha,\mathbf{r}+\mathbf{a}_\alpha/2} + c_{p,\mathbf{r}_p} c_{\alpha,\mathbf{r}-\mathbf{a}_\alpha/2}\right) + h.c.\right]$$

Here $d-p$ PDW order parameter is $\Delta_{pd_\alpha}(\mathbf{r}_p) = \Delta^{pd}_\alpha \cos(\mathbf{M}_\alpha \cdot \mathbf{r}_p)$, where $\mathbf{r}_p$ is the location of the unit cell for the $p_z$ orbital. These real-space pairing modulations are illustrated in the Fig. S3**a** (blue and red lines). By employing the band unfolding, the typical unfolded Bogoliubov Fermi states are displayed in Fig. S3**d**. We find that the central pocket is gapped out by the $s$-wave pairing and sizable gap opening appears at the crossing points between triangular and circular pockets around the M point due to the PDW. This generates Fermi surface segments (tips of the triangular pockets) around the $M$ point and $\Gamma$ (with a weaker intensity). The experimentally detected arc-like scattering features $q_{d1}$ and $q_{d2}$ can be assigned to these Fermi states accordingly.

Secondly, we consider the $d-d$ PDW state and the corresponding Hamiltonian is,

$$\mathcal{H}^{dd}_{\text{PDW}} = \frac{1}{2}\sum_{\mathbf{r}} \sum_{\alpha\beta\gamma \in \{\epsilon_{\alpha\beta\gamma}=1\}} \epsilon_{\alpha\beta\gamma} \Delta_{dd,\alpha}(\mathbf{r})[c_{\beta,\mathbf{r}} c_{\gamma,\mathbf{r}+\mathbf{a}_\alpha/2} - c_{\beta,\mathbf{r}} c_{\gamma,\mathbf{r}-\mathbf{a}_\alpha/2} + h.c.]$$

The PDW order parameter is $\Delta_{dd,\alpha}(\mathbf{r}) = \Delta_\alpha^{dd}\cos(\mathbf{M}_\alpha \cdot \mathbf{r})$, $\Delta_0 = 0.02$ and $\Delta_\alpha^{dd} = 0.01$ (Fig. S3b). The obtained Bogoliubov Fermi states are displayed in Fig. S3e. Compared with the $d-p$ PDW state, the prominent feature is that the Fermi surfaces around the M point is further gapped. Consequently, we could not assign a scattering channel to produce $\boldsymbol{q}_{d2}$ which is inconsistent with our experiments.

Finally, we consider the $p-p$ PDW state and the corresponding Hamiltonian is,

$$\mathcal{H}_{PDW}^{pp} = \sum_{\mathbf{r}} \sum_\alpha \Delta_{pp}(\mathbf{r}_p) \left( c_{p,\mathbf{r}_p} c_{p,\mathbf{r}_p} + h.c. \right)$$

The PDW order parameter is $\Delta_{pp}(\mathbf{r}_p) = \Delta^{pp} \sum_\alpha \cos(\mathbf{M}_\alpha \cdot \mathbf{r}_p)$, $\Delta_0 = 0.02$ and $\Delta^{pp} = 0.005$ and is shown in Fig. S3c. The $p-p$ PDW will not introduce additional gap opening as the $p$-orbital Fermi surface is already gapped by the uniform pairing. The Bogoliubov Fermi states of $d$-orbital are almost continuous and will not generate arc features in the quasi-particle interference spectra, inconsistent with the arc-like quasi-particle interference features in our experiments.

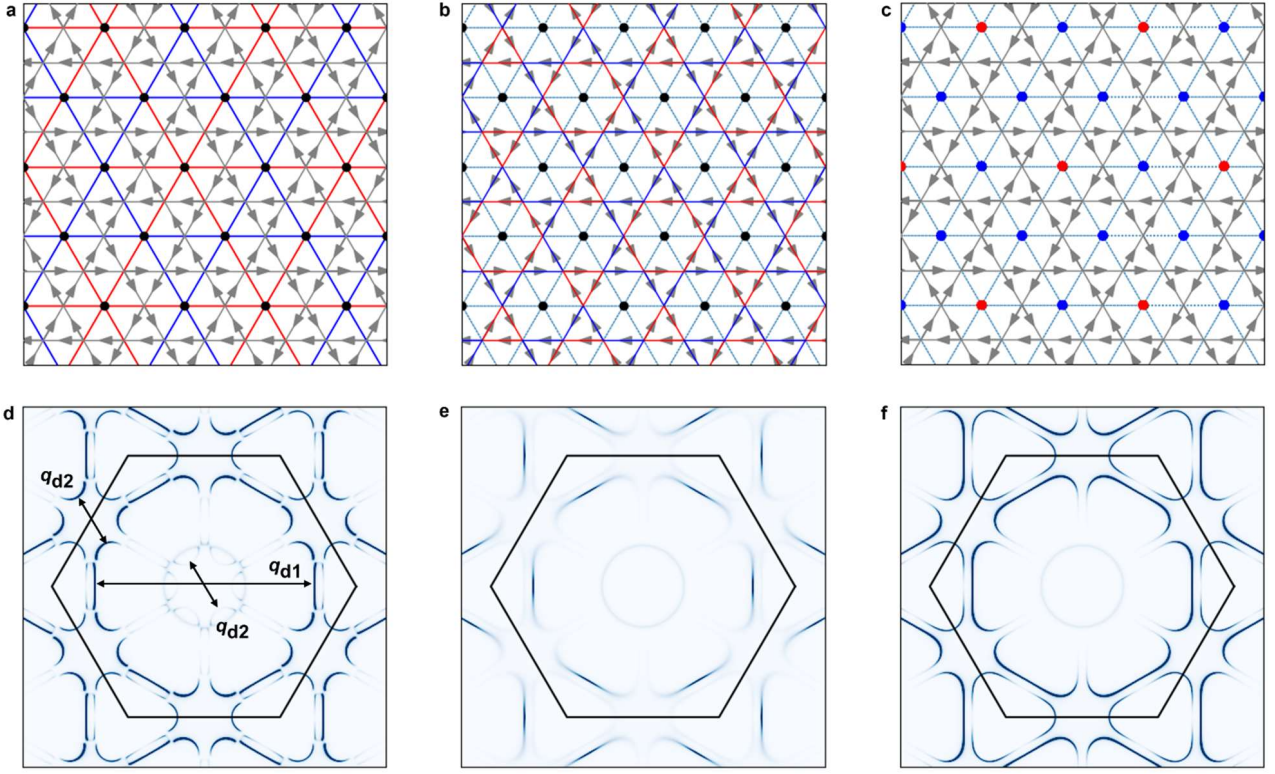

**Figure S3. Different PDW states and corresponding Bogoliubov Fermi surfaces. a-c,** Real space order parameters of $p-d, d-d, p-p$ PDW orders, respectively. The grey arrows in **a-c** indicate loop current order in kagome lattice. Black dots in **a-b** show onsite superconductivity order $\Delta_0 = 0.02$ on $p_z$ orbital. Red and blue line denote the positive and negative PDW pairing. **a** $\Delta_\alpha^{dp} = 0.015$, **b** $\Delta_\alpha^{dd} = 0.01$, **c** $\Delta^{pp} = 0.005$ with red (blue) dots representing strong (weak) pairing. **d-f,** Bogoliubov Fermi surfaces of the $p-d, d-d, p-p$ PDW state, respectively.